\documentclass[aps,prb, preprint,amsmath,amssymb,superscriptaddress]{revtex4-2}
\usepackage{graphicx}
\usepackage{dcolumn}
\usepackage{bm}

\usepackage{xcolor}
\usepackage{amsmath} 
\usepackage{mathrsfs}
\usepackage{graphicx}
\usepackage{dcolumn}
\usepackage{bm}
\usepackage{anyfontsize}
\usepackage[utf8]{inputenc}
\usepackage[T1]{fontenc}
\usepackage{mathptmx}
\usepackage{etoolbox}

\begin{document}


\title[Sample title]{Ultrafast dynamics of carriers, coherent acoustic phonons and strain pulses in BiSbTe$_{1.5}$Se$_{1.5}$ topological insulator thin films}

\author{Anupama Chauhan}
 \affiliation {Department of Physical Sciences, Indian Institute of Science Education and Research Kolkata, Mohanpur, 741246, West Bengal, India.}
\author{Sidhanta Sahu}%
 \affiliation {Department of Physical Sciences, Indian Institute of Science Education and Research Kolkata, Mohanpur, 741246, West Bengal, India.}
 
 \author{ Poulami Ghosh}%
 \affiliation {Department of Physical Sciences, Indian Institute of Science Education and Research Kolkata, Mohanpur, 741246, West Bengal, India.}

\author{Dheerendra Singh}%
 \affiliation {Department of Physical Sciences, Indian Institute of Science Education and Research Kolkata, Mohanpur, 741246, West Bengal, India.}

\author{Sambhu G Nath}%
 \affiliation {Department of Physical Sciences, Indian Institute of Science Education and Research Kolkata, Mohanpur, 741246, West Bengal, India.}

\author{Anjan Kumar N M}%
 \affiliation{Department of Physical Sciences, Indian Institute of Science Education and Research Kolkata, Mohanpur, 741246, West Bengal, India.}

\author{P.K. Panigrahi}%
 \affiliation {Department of Physical Sciences, Indian Institute of Science Education and Research Kolkata, Mohanpur, 741246, West Bengal, India.}
 \affiliation {Centre for Quantum Science and Technology, Siksha ‘O’ Anusandhan university, Bhubaneswar, 751030, Odisha, India.}
 
 \author{Chiranjib Mitra}%
 \affiliation {Department of Physical Sciences, Indian Institute of Science Education and Research Kolkata, Mohanpur, 741246, West Bengal, India.}
 
\author{N. Kamaraju}%
\email{nkamaraju@iiserkol.ac.in} 
\altaffiliation {Department of Physical Sciences, Indian Institute of Science Education and Research Kolkata, Mohanpur, 741246, West Bengal, India.}

\date{\today}
 
\begin{abstract}

We investigate the ultrafast carrier, coherent acoustic phonons (CAPs), and acoustic strain pulse dynamics in topological insulator BiSbTe$_{1.5}$Se$_{1.5}$ (BSTS) thin films of varying thickness using degenerate pump-probe reflection spectroscopy. Here, Sapphire has been chosen as the main substrate due to its maximum acoustic reflectivity at the BSTS-sapphire interface compared to BSTS-GaAs, BSTS-Si, and BSTS-MgO interfaces. For the films with thickness more than twice the penetration depth, the transient reflectivity data predominantly exhibits travelling acoustic strain pulses (TASP) on the top of single-exponential electronic decay ($\sim$ 2 ps). In contrast, films with thickness less than penetration depth are dominated by CAPs and a bi-exponential electronic background with decay constants $\tau_{1}$ $\sim$ 2 ps and $\tau_{2}$ $\sim$ 260-380 ps. The observed TASP dynamics are well-described by a theoretical acoustic strain model. Further, to elucidate the underlying physical mechanisms governing the behaviour of photo-excited carriers, CAPs, and strain pulses, we performed carrier density- and temperature-dependent (7–294 K) studies on BSTS films with thicknesses of 22 nm and 192 nm. In the 22 nm film, the two decay constants $\tau_{1}$ and $\tau_{2}$ increase with carrier density at room temperature but decrease with temperature at a carrier density of 1.7 $\times$ 10$^{19}$ cm$^{-3}$. A detailed analysis suggests that $\tau_{1}$ arises from electron-phonon scattering and carrier diffusion, while $\tau_{2}$ likely results from defect-assisted and phonon-assisted recombination. Furthermore, increasing the sample temperature leads to anharmonic decay induced softening of $\sim$ 14 $\%$ in the phonon frequency and an anomalous $\sim$ 48 $\%$ decrease in the phonon damping parameter due to reduced Dirac surface electron and acoustic phonon scattering. Temperature-dependent studies of the 192 nm film reveal a $\sim$ 7 $\%$ reduction in sound velocity, in contrast to the 14 $\%$ reduction observed in the 22 nm film.

\end{abstract}

\maketitle

\section{Introduction}

Topological insulator (TI) thin films have emerged as a fascinating class of materials in condensed matter physics due to their remarkable ability to support conducting surface states while remaining insulating in the bulk \cite{kane2006new,moore2009next,moore2010birth}. 
These surface states arise from an interplay of spin-orbit coupling and time-reversal symmetry, which induces a band inversion, leading to the crossing of conduction and valence bands at the surface, resulting in symmetry protected gapless Dirac surface states \cite{roushan2009d,hasan2010colloquium}. Many well-known TI materials, such as Bi$_{2}$Se$_{3}$ and Bi$_{2}$Te$_{3}$, suffer from poor bulk insulating properties, limiting the ability to investigate the anticipated novel surface phenomena \cite{hasan2010colloquium,qu2010quantum}. In contrast, BiSbTe$_{1.5}$Se$_{1.5}$ (BSTS) has emerged as the promising TI material displaying robust insulating bulk states \cite{di2012optical,singh2017linear,gopal2017topological}, making it ideal for exploring the exotic Dirac surface states. BSTS demonstrates higher surface conductivity \cite{taskin2011observation,xia2013indications} and lower thermal conductivity \cite{pathak2022strong} compared to Bi$_{2}$Se$_{3}$, Bi$_{2}$Te$_{3}$, and Bi$_{2}$Te$_{2}$Se, making it an attractive candidate for applications in quantum computing \cite{sarma2015majorana,kitaev2003fault}, spintronics \cite{jamali2015giant, kondou2016fermi}, and thermoelectrics \cite{xu2017topological,baldomir2019behind,tan2013improved,article}.

Ultrafast studies are essential for advancing thermoelectric devices, as they provide critical insights into the fundamental processes governing charge and heat transport on ultrafast timescales. These ultrafast pump-probe studies have been utilised to infer the non-equilibrium surface  and bulk electron, and phonon dynamics, which are crucial for optimizing thermoelectric materials \cite{thomsen1986surface,thomsen1984coherent,wang2010acoustic,cheng2014temperature,fonseca2024picosecond,lai2014temperature}. In these experiments, upon irradiation of a femtosecond pump pulse on a material, acoustic phonons along with the photoexcited carriers are generated. A time delayed probe pulse is then used for detection, which gets reflected from the top surface of the material and also from the travelling acoustic phonons. This in turn results in the interference between these two reflections at the detector, revealing rich information about the acoustic phonons in the material. The mechanism for the generation of these acoustic phonons can involve the contribution of thermo-elastic stress \cite{thomsen1986surface,wu2007femtosecond,ruello2009laser} and deformation potential stress \cite{thomsen1986surface,wu2007femtosecond,ruello2009laser,lim2005coherent,wang2005propagating} in semiconductors.

The propagation of these acoustic phonons in thin films depends on different factors like the thickness of the film, excited carrier density and sample temperature. For instance, if the thickness of the sample is much more than the penetration depth ($\xi$), it can host travelling acoustic strain pulses (TASP) \cite{thomsen1986surface,fonseca2024picosecond}. Conversely, when the thickness is less than the penetration depth, the film can act as an acoustic phonon resonator (APR) \cite{glinka2015acoustic}. To the best of our knowledge, there is no comprehensive investigation of how the TASP emerges from the CAPs. Thus, investigating the photoexcited carriers and propagation of acoustic phonons in BSTS thin films is expected to provide valuable insights, which are essential for understanding thermal transport in TI-based thermoelectric devices.  In this regard, it is important to inspect if there is any coupling between the acoustic phonons and the surface electrons as the energy of the acoustic phonons is comparable to surface electrons near the Dirac point. 

In this study, we have presented an extensive investigation of the carrier and CAP dynamics in BSTS thin films of various thicknesses, as a function of carrier density and temperature, using degenerate pump-probe reflection spectroscopy. The time-resolved reflectivity signal reveals the presence of acoustic strain pulses, along with an exponentially decaying response from the excited carriers. The BSTS-Sapphire interface is found to have the best acoustic strain pulse reflectivity of $\sim$ 42 $\% $ in comparison to the interface of BSTS-GaAs, BSTS-Si and BSTS-MgO. Hence, the BSTS thin film with sapphire interface has been selected to study the influence of varying thickness (d) on the transient reflectivity data. Our experiments unveil the interplay between two distinct regimes, where the dynamics of acoustic phonons are predominantly influenced by TASP for higher thickness of the film and APR modes as the film thickness decreases. The transient reflectivity data for the films with d $\gtrsim$ 2$\xi$ mainly contains the TASP on top of a single-exponential electronic background with a decay time constant of $\sim$ 2 ps. These TASPs are simulated by the well-known theoretical acoustic strain model \cite{thomsen1984coherent,thomsen1986surface}. And the transient reflectivity data for BSTS films with d < $\xi$ is dominated by APR modes and a bi-exponential electronic background. To further understand the behaviour of carriers and CAP dynamics, a detailed carrier density and temperature dependent study is carried out on 22 nm and 192 nm thick BSTS films on sapphire. The bi-exponential electronic background for the 22 nm film ($\tau_{1}$ $\sim$ 2 ps and $\tau_{2}$ $\sim$ 350-450 ps) and CAP parts display strong dependence on the carrier density and temperature. Both the $\tau_{1}$ and $\tau_{2}$ are observed to increase  by $\sim$ 12 $\%$ and $\sim$ 20 $\%$ respectively with the carrier density variation from 0.8 $\times$ 10$^{19}$ - 13.5 $\times$ 10$^{19}$ cm$^{-3}$. And both $\tau_{1}$ and $\tau_{2}$ decrease by $\sim$ 15 $\%$ and $\sim$ 12 $\%$ respectively as the sample is heated from 7-294 K. A detailed analysis of $\tau_{1}$ and $\tau_{2}$ as functions of carrier density (at room temperature) and temperature (at constant carrier density) reveals that $\tau_{1}$ arises from the interplay between electron-phonon scattering and carrier diffusion, while $\tau_{2}$ is likely to originate from the combined effects of defect-assisted and phonon-assisted recombination. The acoustic phonon lifetime and frequency for the 22 nm film, decreases with increasing carrier density, suggesting an enhancement in electron-phonon scattering. The acoustic phonon frequency softens by $\sim$ 14 $\%$ with increasing temperature, in line with normal behavior, while the phonon damping factor shows an anomalous decrease as the temperature rises due to decreased scattering between the Dirac surface electrons and acoustic phonons. Temperature-dependent studies reveal a $\sim$ 7 $\%$ reduction in sound velocity for the 192 nm film, in contrast to $\sim$ 14 $\%$ reduction observed in the 22 nm film as the sample temperature is increased.

\section{Methods}

BSTS crystallizes in a rhombohedral structure with R$\bar{3}$m space-group symmetry. In this work, pulsed laser deposition (PLD) technique has been utilised to deposit all BSTS thin films on various substrates of sapphire (Al$_{2}$O$_{3}$), GaAs, MgO and Si. The details of the sample growth procedure using the PLD technique are discussed in the Supplemental Material \cite{Supplemental}, Sec. I and the details of the substrates are summarized in supplemental material \cite{Supplemental}, table TS1. The characterization of the thin films using X-ray diffraction (XRD), Raman spectroscopy and scanning electron microscopy (SEM) is discussed in the supplemental material \cite{Supplemental}, Fig. S1. The thickness of the films is measured using cross-sectional SEM technique (see supplemental material \cite{Supplemental}, Fig. S2). 

\begin{figure*}
\includegraphics[width=0.65\linewidth]{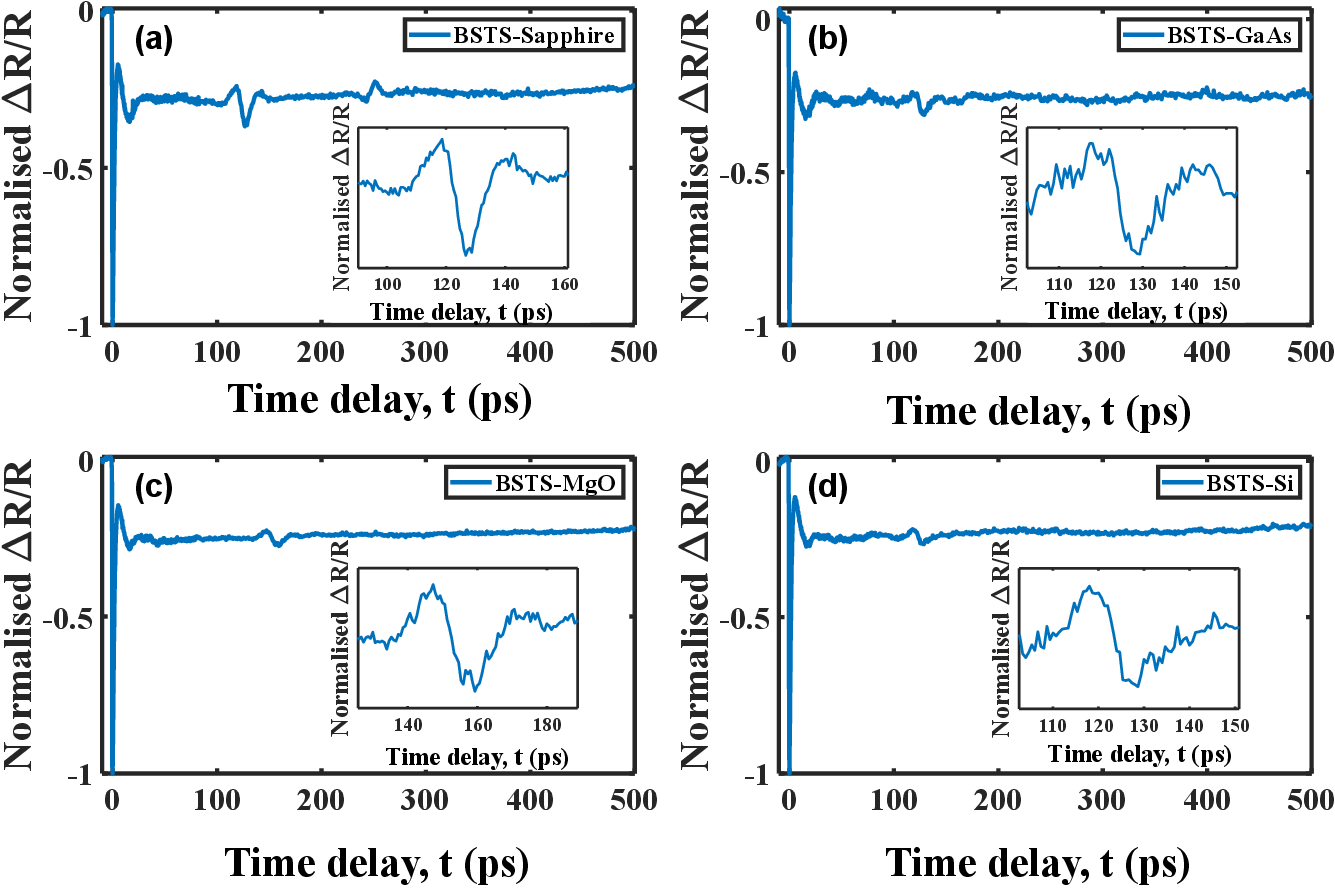}
\caption{ Time-resolved normalised $\Delta$R/R signal as a function of time delay between the pump and probe showing acoustic strain pulses on the top of exponentially decaying electronic background observed for (a) BSTS (164.9 nm $\pm$ 2.5 nm) - Sapphire interface, (b) BSTS (170.5 nm $\pm$ 2.6 nm) - GaAs interface, (c) BSTS (211.0 nm $\pm$ 3.1 nm) - MgO interface and (d) BSTS (168.7 nm $\pm$ 2.6 nm) - Si interface  respectively. Inset in each figure shows the enlarged view of first echo of the acoustic strain pulse. Here, the pump and probe fluence are kept constant at 87 $\mu$J/cm$^{2}$ and 6 $\mu$J/cm$^{2}$ respectively.} \label{fig_1}
\end{figure*}

The time-resolved reflectivity experiments have been performed using a home-built reflection geometry pump-probe spectroscopy setup (see supplemental \cite{Supplemental}, Fig. S3). Our setup uses pulsed laser output of an amplifier (RegA 9050, Coherent Inc.) operated at 250 kHz with pulse width of $\sim$ 60 fs and central photon energy of $\sim$ 1.57 eV. The sample temperature is varied using a closed-cycle helium cryostat (OptistatDryBLV, Oxford Instruments). Further details of the pump-probe setup and measurements are provided in the supplemental material \cite{Supplemental}, sec. II.

For fluence dependent studies, the pump fluence is varied from 11 $\mu$J/cm$^{2}$ to 174 $\mu$J/cm$^{2}$ which corresponds to the carrier density of $\sim$ 0.8 $\times$ $10^{19}$ cm$^{-3}$ to 13.5 $\times$ $10^{19}$ cm$^{-3}$. The probe fluence is kept constant at 6 $\mu$J/cm$^{2}$. The pump fluence is chosen such that the sample remains undamaged. For the analysis, time-resolved $\Delta$R/R signal is normalised with respect to its peak value.

\section{Results and Discussion}
\subsection{Effect of Film thickness on acoustic phonons}

The transient reflectivity ($\Delta$R/R) data for BSTS films on four substrates of Sapphire, GaAs, MgO and Si, are shown in Fig. \ref{fig_1}(a)-(d). Upon irradiation with the pump pulse, the excitation of non-equilibrium electrons leads to a negative dip ($\Delta$R/R) at zero time delay (see schematic Fig. \ref{fig_6}(a)). These photoexcited carriers then undergo carrier-phonon scattering lasting $\sim$ few ps (see schematic Fig. \ref{fig_6}(b)) and subsequently undergo interband recombination with the timescales of $\sim$ hundreds of ps as the signal exponentially recovers towards zero \cite{prakash2017origin}. On top of this electronic background, the acoustic strain pulses in the GHz frequency range are observed. These acoustic strain pulses are generated due to the thermo-elastic \cite{thomsen1984coherent,thomsen1986surface} and deformation potential stress \cite{wu2007femtosecond} (the theory of generation and detection of acoustic strain pulses is discussed in the supplemental material \cite{Supplemental}, sec. III). It can be seen from Fig. \ref{fig_1} that, on top of the main acoustic strain pulse, there are also echo pulses that occur due to the multiple reflections of the primary strain pulse. 

\begin{figure*}
\includegraphics[width=1.0\linewidth]{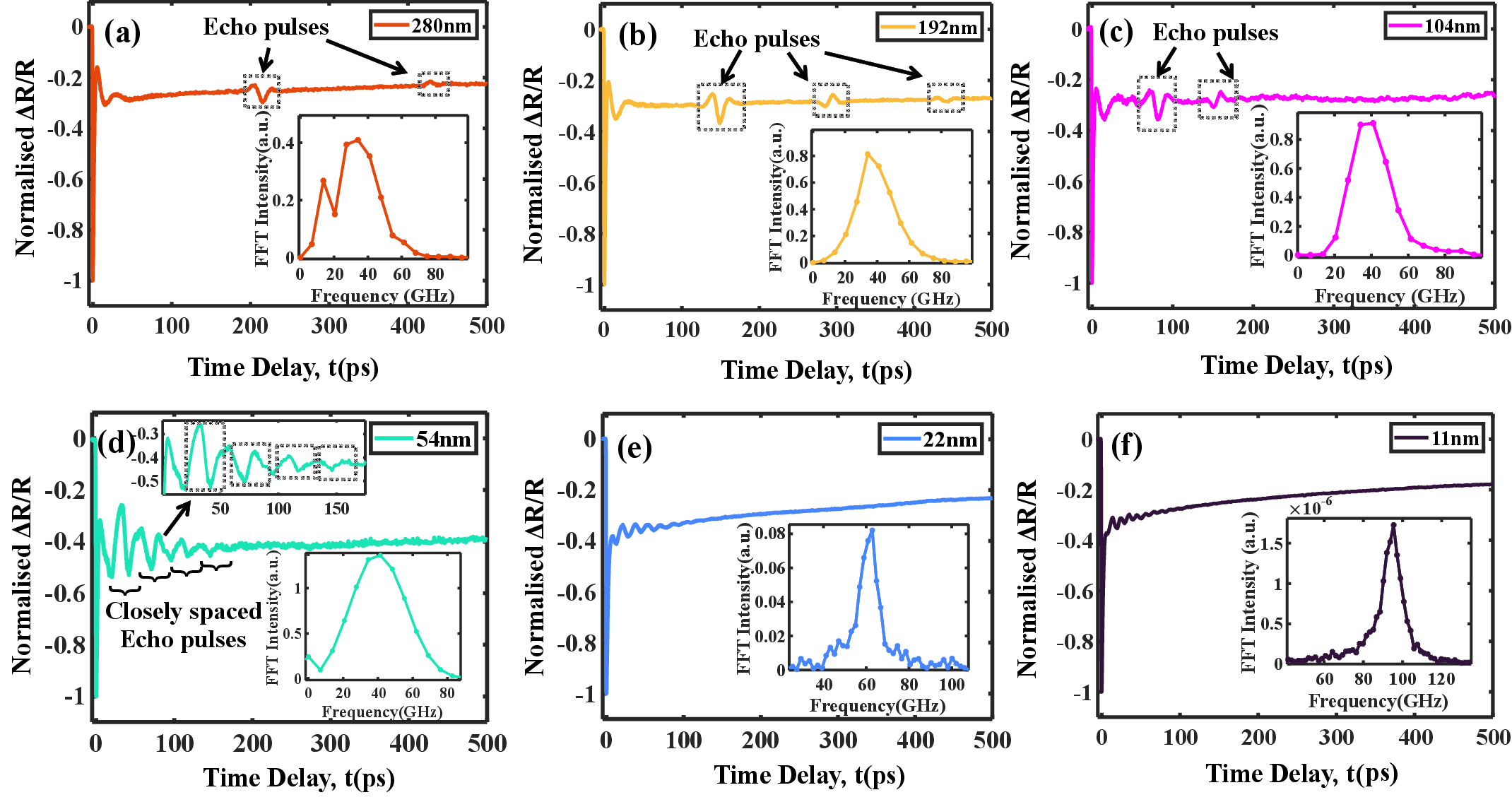}
\caption{ Time-resolved normalised $\Delta$R/R signal observed for the different thicknesses of BSTS thin film deposited on Sapphire substrate. (a) BSTS sample of thickness 280 nm shows clearly separated acoustic strain pulses with first and second echo appearing at 215 ps and 428 ps respectively. (b) For 192 nm thick BSTS, the first, second and third echo appears at 148 ps, 295 ps and 442 ps respectively. (c) For 104 nm thick film, first and second echo appears at 81 ps and 160 ps respectively. (d) For 54 nm thick film, four closely spaced echoes, at 32 ps, 71 ps, 109 ps and 146 ps are observed. For (a)-(d) the inset shows the FFT spectra of the extracted first echo of the acoustic strain pulse. (e) The data for the 22 nm thick film displays the oscillatory motion with a period of 16 ps on the top of the bi-exponential decay. Inset shows the FFT of the extracted oscillation peaking at 61 GHz. (f) The data for the 11 nm film shows the oscillatory motion with a period of 10.5 ps in addition to the bi-exponential decay. Inset shows the FFT of the extracted oscillation peaking at 95 GHz. Here, the pump and probe fluence are kept constant at 87 $\mu$J/cm$^{2}$ and 6 $\mu$J/cm$^{2}$ respectively. }  \label{fig_2}
\end{figure*}

The calculated values (see supplemental material \cite{Supplemental}, Table TS3 for details) of acoustic reflection coefficient (r$_{a}$) for these pulses is found to be maximum for BSTS-Sapphire interface and minimum for BSTS-Silicon interface as the acoustic strain pulse echoes have maximum amplitude for BSTS-Sapphire interface. In addition, it is seen that the successive echoes are inverted with respect to each other (Fig. \ref{fig_1}(a)). This can be understood by considering the boundary conditions at the two interfaces of the film, namely the BSTS-Air and BSTS-Substrate interfaces. At the BSTS-Air interface, a phase shift of ‘$\pi$’ occurs because the acoustic impedance of air is lower than that of the BSTS film. In contrast, at the BSTS-Substrate interface, no phase shift is observed, as the acoustic impedance of the substrate is higher than that of the BSTS film \cite{thomsen1986surface}.

Since BSTS-Sapphire interface has the best acoustic reflection coefficient, further time-resolved investigations are carried out on BSTS films of varying thickness on the sapphire substrate. It should be noted here that the effective wavelength of the acoustic strain pulse is $\sim$ 2$\xi$ \cite{thomsen1986surface} ($\xi$ $\sim$ 26 nm is calculated from UV-visible measurements which is in agreement with the previous literature \cite{cheng2014temperature}). The thickness of the BSTS film is thus chosen to be varied from below and above the optical penetration depth i.e. from 11 nm to 280 nm.  For d $\geq$ 2$\xi$ ($\gtrsim$ 52 nm i.e. for 104 nm, 192 nm and 280 nm films), the clearly separated acoustic strain pulses on top of the single-exponential electronic background ($\sim$ 2 ps) are observed in the time domain (Fig.\ref{fig_2}(a)-(d)) with separation 2d/v$_{s}$, where v$_{s}$ denotes the longitudinal sound velocity. And the Fast Fourier transform (FFT) of the extracted first echo of the acoustic strain pulse (see supplemental material \cite{Supplemental},  sec. V for details) shows a broad spectrum peaking around 35-40 GHz (insets in Fig. \ref{fig_2}(a)-(d)). For the 54 nm film, with d $\sim$ 2$\xi$, the strain echo pulses are very closely spaced (shown in Fig. \ref{fig_2}(d)).  As the thickness of the film is further reduced (< 52 nm i.e. for 11 nm and 22 nm films), the pump beam penetrates into the substrate completely and the entire film gets excited. Here, we observe clear oscillations in time-domain with narrower FFT intensity spectrum (Fig. \ref{fig_2}(e), (f)). This is because a thin film with thickness less than the penetration depth acts as an acoustic phonon cavity whose fundamental resonance frequency is solely determined by its thickness \cite{glinka2015acoustic}. Thus, only the acoustic phonon whose frequency matches with the fundamental cavity resonance survives, displaying spectral narrowing, and these are referred as APR modes. The relationship between the excited fundamental resonant mode frequency ($\nu_{LA}$) and thickness in thin films is given by, $\nu_{LA}$= v$_{s}$/2d. The frequency of the excited modes obtained after subtracting the electronic background (see supplemental \cite{Supplemental}, sec. V for details) for the 11 nm and 22 nm films is $\sim$  95 GHz and $\sim$ 61 GHz respectively (shown in the inset of Fig. \ref{fig_2}(e), 2(f) respectively). Here, the electronic background is of bi-exponential nature ($\tau_{1}$ $\sim$\ 2.0-2.6 ps and $\tau_{2}$ $\sim$\ 263-380 ps) and dominant as compared to the data for d $\geq$ 2$\xi$ in the pump probe time delay (see supplemental \cite{Supplemental}, Fig. S7).

\begin{figure}
\includegraphics[width=1\linewidth]{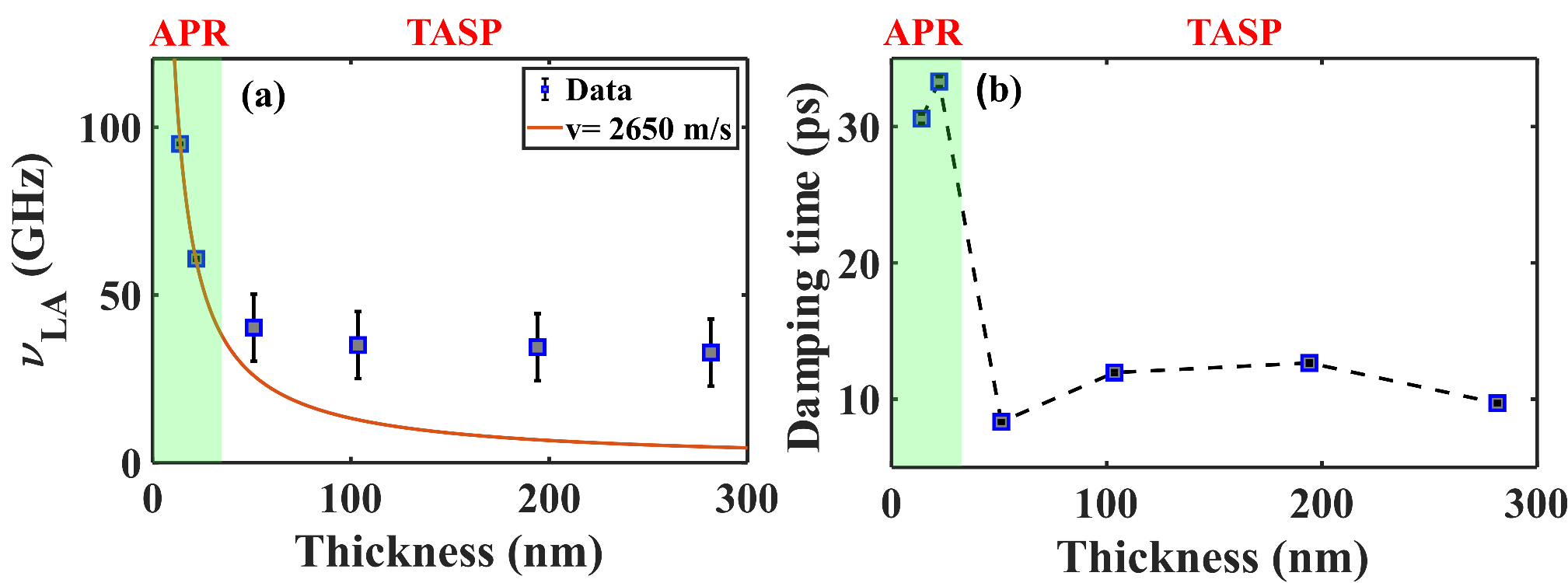}
\caption{ (a) The variation of central frequency of FFT intensity for the CAPs as a function of thickness. Orange solid line shows the plot with $\nu_{LA}$= 2d/v$_{s}$ where v$_{s}$= 2650 m/s. (b) The damping time ($\frac{1}{\pi* w}$, where "$w$" is the full width at half maximum of the FFT intensity spectrum) of the CAPs with increasing thickness. Here, the data are represented by blue open squares and dotted lines are guide to the eye.}\label{fig_3}
\end{figure}

The variation of the acoustic phonon frequency (central frequency of the FFT intensity spectra) with increasing thickness is shown in Fig. \ref{fig_3}(a), and is expected to follow the expression $\nu_{LA}$ = v$_{s}$/2d \cite{glinka2015acoustic}(orange line plotted using v$_{s}$ = 2650 m/s). A deviation from this expression is observed for d > 26 nm, as these correspond to the central frequencies of the FFT intensity spectrum of TASP, rather than the CAP frequency. Fig. \ref{fig_3}(b) shows the variation of the damping time \cite{kamaraju2010large} (given by $\frac{1}{\pi*w}$ where "$w$" is the full width at half maximum of the FFT intensity spectrum) of the acoustic phonons with thickness. Note that the lifetime of the acoustic phonons decreases drastically as we go from APR to TASP regime of thickness due to the formation of the strain pulses.

\begin{figure*}
\includegraphics[width=0.8\linewidth]{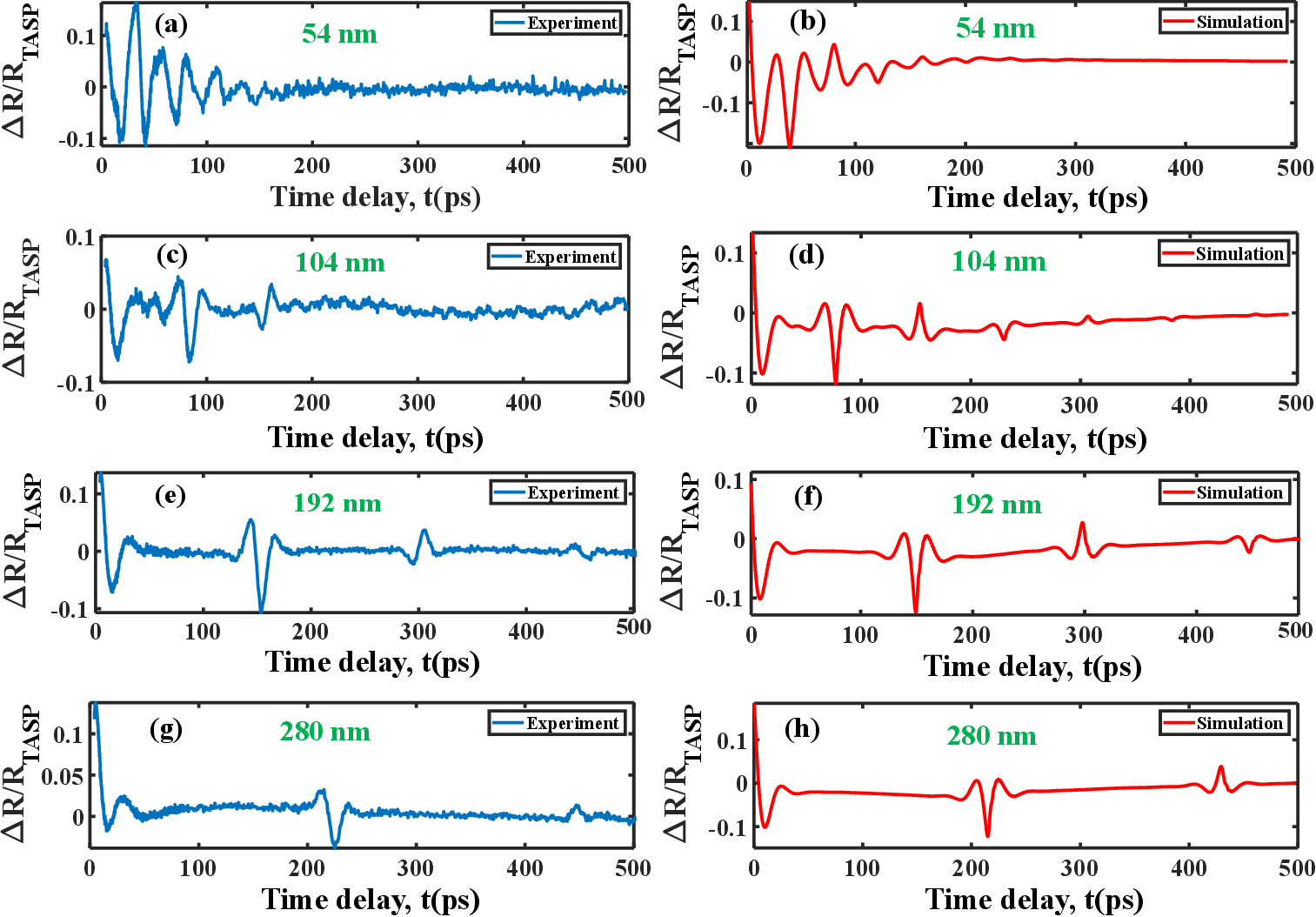}
\caption{ Simulation of the acoustic strain pulses using Thomsen model (\cite{thomsen1986surface,wu2007femtosecond}). Blue curves (a), (c), (e) and (g) are the extracted $\Delta$R/R$_{TASP}$ signal for the 54 nm, 104 nm, 192 nm and 280 nm BSTS film respectively. And red curves (b), (d), (f) and (h) are the simulations using the Thomsen model (\cite{thomsen1986surface,wu2007femtosecond}) for the 54 nm, 104 nm, 192 nm and 280 nm BSTS film respectively.} \label{fig_4}
\end{figure*}

Taking into account that the origin of the acoustic strain pulses/CAPs in BSTS is linked to the thermo-elastic stress \cite{thomsen1986surface,hase2015coherent} (see supplemental material \cite{Supplemental}, sec. III for the details of the theory), the complete acoustic strain model proposed by Thomsen et al. \cite{thomsen1986surface} has been used to simulate the experimental data. Fig 4. shows the experimental and simulated data for 54 nm (see Fig. \ref{fig_4}(a)-(b)), 104 nm (see Fig. \ref{fig_4}(c)-(d)), 192 nm (see Fig. \ref{fig_4}(e)-(f)), and 280 nm (see Fig. \ref{fig_4}(g)-(h)) films. Here, a clear alignment between theoretical and experimental results is evident. 

Now, using the calculated sound velocity of 2650 $\pm$ 36 m/s in the BSTS film and the observed period of oscillation (T) for CAPs (for d < $\xi$) or the time separation (T$'$= t$_{2}$- t$_{1}$) between the acoustic strain pulses (for d $\gtrsim$ $\xi$), the thickness of the films can be estimated and these values are found to be in close agreement with the thickness measured from cross-sectional SEM (see Table I and the supplemental material \cite{Supplemental}, Fig. S2). 

\begin{table}
\caption{\label{tab:table2}Estimated thickness and measured thickness (from cross-sectional SEM) of all the films}
\begin{ruledtabular}
\begin{tabular}{cccc}
Period of Osc.& Est. Thickness& Measured thickness from \\ T (ps) & d = v$_{s}$T/2 (nm)  & cross-sectional SEM (nm) \\
\hline
10.5$\pm$ 0.2   &	13.9$\pm$ 0.5	&11.4$\pm$2.0\\
16.6  $\pm$ 0.2   &	22.1$\pm$ 0.6	&22.5$\pm$2.0\\
\hline
Time sep. b/w echoes& Est. Thickness&  Measured thickness from \\ T$'$= $\tau_{2}-\tau_{1}$ (ps) &d = v$_{s}$T$'$/2 (nm) & cross-sectional SEM (nm)\\
\hline

38.5 $\pm$ 0.2 & 51.0$\pm$ 0.6  &53.9 $\pm$2.0\\
78.2   $\pm$ 0.2 & 103.6$\pm$ 1.7 &104.3 $\pm$2.0\\
146.5  $\pm$ 0.2 & 194.1$\pm$ 2.9 &191.9 $\pm$2.0\\
212.4  $\pm$ 0.2 & 281.5$\pm$ 4.1 &280.2 $\pm$2.0\\

\end{tabular}
\end{ruledtabular}
\end{table}

\subsection{Effect of photoexcited carriers on CAPs and strain dynamics}

\begin{figure}
    \centering
    \includegraphics[width=0.85\linewidth]{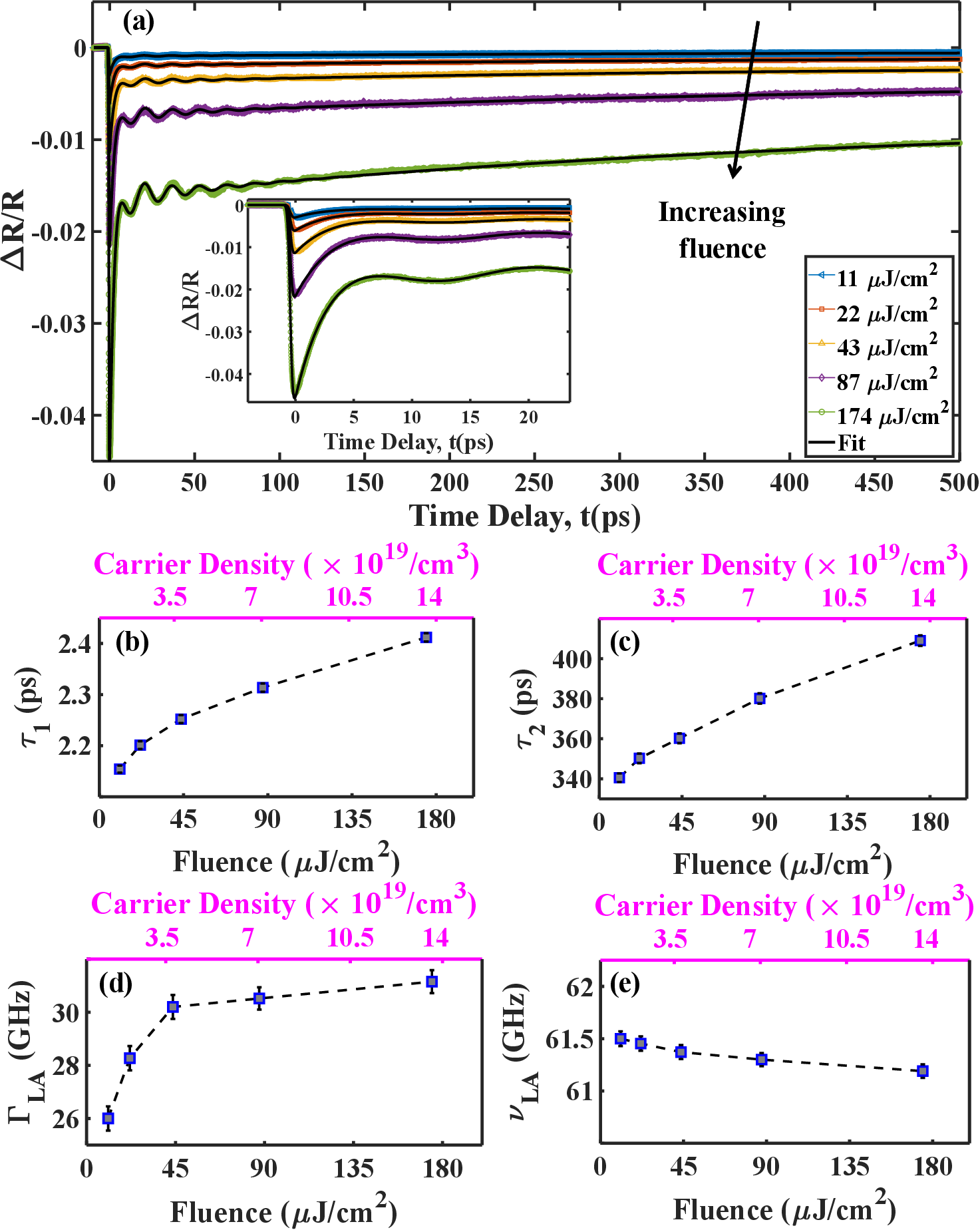}
    \caption{(a) The time-resolved $\Delta$R/R signal at various pump fluences from 11 $\mu$J/cm$^{2}$ to 174 $\mu$J/cm$^{2}$ (different colour symbols shows different fluences) for 22 nm BSTS film at room temperature and black solid line shows the fit using eq. (\ref{eq_1}). The direction of the arrow indicates that the pump fluence is increasing from top curve to the bottom curve. (b) and (c) display the variation of $\tau_{1}$ and $\tau_{2}$ respectively. (d) and (e) show the behaviour of phonon damping parameter $\Gamma_{LA}$ ($1/\tau_{LA}$) and frequency, $\nu_{LA}$ of the CAP respectively. Here, the data are represented by blue open squares and dotted lines are guide to the eye.} \label{fig_5}     
\end{figure}

 As the electron and CAPs dynamics are more prominent for d < $\xi$, we selected the BSTS sample with d = 22 nm for further investigation by varying the carrier density at room temperature (see Fig. \ref{fig_5}(a)). The $\Delta$R/R data show an oscillatory pattern due to the CAPs on top of the bi-exponential electronic background. The data could be well fitted (for $\mathscr{R}^{2}$ values, see supplemental material \cite{Supplemental}, Table TS5) with the following function,

\begin{equation} 
    \begin{split}
        \frac{\Delta R}{R}= \bigl[\bigl\{A_{0}+\sum_{i=1}^{2} A_{i} e^{-t/\tau_{i}} + B_{LA} e^{-t/\tau_{LA}} cos (2\pi \nu_{LA} t + \phi_{LA})\bigr\} \\
        \times (1-e^{-t/\tau_{r}})\bigr] \otimes CC(t)
    \end{split}\label{eq_1}
\end{equation}
  
 where the multiplication factor, ($1-e^{-t/\tau_{r}}$) of eq. (\ref{eq_1}) represents the initial rise of the $\Delta$R/R signal giving a rise time, $\tau_{r}$ $\sim$ 255 fs, that corresponds to the thermalisation of excited carriers \cite{prakash2017origin}. CC(t) here is the pump-probe cross correlation function with FWHM of $\sim$ 235 fs. The first term inside the parenthesis,  $A_{0}$ represents amplitude of the long-lived carrier recombination dynamics. The second term in the parenthesis of eq. (\ref{eq_1}) signifies the bi-exponential electronic background ($A_{1}$, $A_{2}$ and $\tau_{1}$, $\tau_{2}$ are the decay coefficients and the decay time constants of the excited carriers respectively). Finally, the third term of the eq. (\ref{eq_1}) inside the parenthesis, is the damped harmonic oscillator function for the CAPs (B$_{LA}$, $\tau_{LA}$, $\nu_{LA}$ and $\phi_{LA}$ are the amplitude, lifetime, frequency and phase of the CAPs).

\begin{figure}
    \centering
    \includegraphics[width=1.0\linewidth]{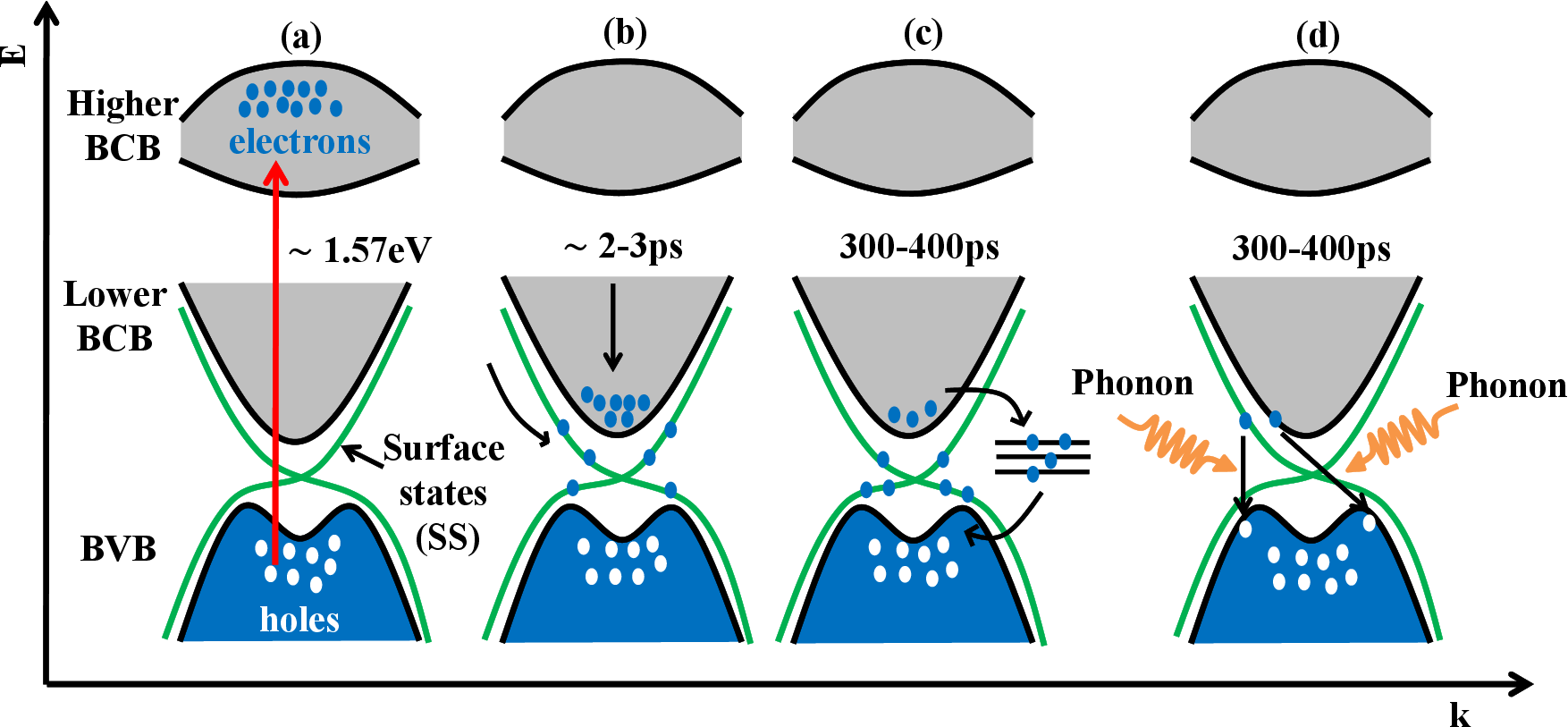}
    \caption{Schematic for the various processes in BSTS. (a) The initial excitation of the electrons upon irradiation with the pump pulse from bulk valence band (BVB) to higher conduction band (BCB). (b) The relaxation of electrons towards the lower BCB edge through electron-phonon scattering and diffusion of carriers. (c) The recombination of electrons and holes via defect states. (d) The recombination process of surface or bulk electrons and holes via phonons.}\label{fig_6}
      
\end{figure}

 The behaviour of the decay time constants at various carrier densities is shown in Fig. \ref{fig_5}(b)-(c). We observe that $A_{1}$ contributes dominantly $\sim$ 70-78 $\%$ of the decay process whereas $A_{0}$ and $A_{2}$ contributes around $\sim$ 12-14 $\%$ and $\sim$ 11-15 $\%$ respectively (see supplemental material \cite{Supplemental}, Fig. S9 for details). The decay time constants, $\tau_{1}$ $\sim$ 2 ps and $\tau_{2}$ $\sim$ 340-410 ps, increase by $\sim$ 12 $\%$ and $\sim$ 20 $\%$ respectively, with the carrier density variation from $\sim$ 0.8 $\times$ $10^{19}$ cm$^{-3}$ to 13.5 $\times$ $10^{19}$ cm$^{-3}$. The fast decay constant $\tau_{1}$, can arise from the thermalisation of carriers via electron-phonon scattering \cite{onishi2015ultrafast,cheng2014temperature,sobota2012ultrafast} and diffusion of the carriers \cite{wu2008ultrafast,PhysRevMaterials.7.054601,kamaraju2010large} (see schematic Fig. \ref{fig_6}(b)). To understand the observed behaviour of $\tau_{1}$ with carrier density, we attempted to fit the data with two-temperature model (TTM) \cite{sobota2012ultrafast,groeneveld1995femtosecond} but found that it did not provide a satisfactory fit. However, a modified version of the TTM is able to capture the trend of $\tau_{1}$ with carrier density (for more details, refer to supplemental material \cite{Supplemental}, sec. VI). Additionally, previous studies \cite{wu2008ultrafast, kamaraju2010large} on tellurium and Bi$_{2}$Te$_{3}$ show that carrier diffusion occurs within $\sim$ 1-2 ps in topological insulators. Thus, by using the diffusion length of 26 nm ($\sim$ $\xi$) and the observed decay time $\tau_{1}$, here in this study, the diffusion constant, D is estimated to be $\sim$ 2.2 cm$^{2}$/s, that is of the similar order as observed in Bi$_{2}$Se$_{3}$ \cite{kumar2011spatially,onishi2015ultrafast} at 300 K (D = 1.2 cm$^{2}$/s). This implies that both electron-phonon scattering and diffusion of carriers are plausibly contributing to the decay time constant $\tau_{1}$. The slow decay time constant $\tau_{2}$ is increasing with the carrier density by $\sim$ 20 $\%$ (340-410 ps). Such a behaviour indicates that $\tau_{2}$ likely originates from the  defect-assisted recombination as the BSTS thin films are known to host Bi, Sb anti-site defects and Se vacancy defects \cite{pandey2019pulsed} (see schematic Fig. \ref{fig_6}(c)). The observed behaviour for $\tau_{2}$ can be explained as follows: as the carrier density increases, the number of available defect states reduces. This reduction leads to a delay in the ability of new carriers to occupy the available defect states, thereby increasing $\tau_{2}$ with rise in carrier density.

Next, we turn our attention to the variation of amplitude (B$_{LA}$), damping parameter ($\Gamma_{LA}$) and frequency ($\nu_{LA}$) of CAPs (see Fig. \ref{fig_5}(d)-(e)). The amplitude of the phonon, B$_{LA}$ (see supplemental material \cite{Supplemental}, Fig. S9(c)), initially increases and then saturates at higher carrier density (which is also seen in bulk CdSe \cite{wu2014ultrafast}). The observed increase in B$_{LA}$ with carrier density further corroborates that the generation of CAPs is resulting from thermo-elastic and deformation potential stresses \cite{thomsen1986surface,wu2014ultrafast}. An increase in carrier density induces greater strain at the surface, which in turn leads to a higher amplitude of the generated phonons. And the saturation of the B$_{LA}$ at high carrier densities may be attributed to an increase in the optical band gap resulting in the saturation of absorption due to the Burstein-Moss effect \cite{zhao2024burstein,dubroka2017interband}. Subsequently, the $\Gamma_{LA}$ (1/$\tau_{LA}$) of CAPs increases by $\sim$ 19 $\%$ with increase in carrier density (see Fig. \ref{fig_5}(d)). This is attributed to the increased scattering interactions between electrons and phonons, which in turn reduces the phonon lifetime. Further, the frequency of the CAPs remains nearly constant with increase in carrier density (see Fig. \ref{fig_5}(e)). This can be understood as follows: The CAPs are generated near the surface of the BSTS film, but they rapidly propagate away from the region of carrier excitation. The penetration depth in the BSTS film is about 26 nm, with the CAPs traversing this thickness in only 9 ps, a time shorter than the single oscillation period ($\sim$ 16 ps). Consequently, the impact of pump fluence on the CAPs' frequency is very less (which is also seen in bulk CdSe \cite{wu2014ultrafast}). Our photoexcited carrier density dependent studies on 192 nm thick samples do not show interesting features except for the monotonic linear increase in its TASP amplitude with increase in the carrier density (data not shown).

\subsection{Effect of temperature on carriers and CAPs dynamics}

\begin{figure}
\includegraphics[width=0.85\linewidth]{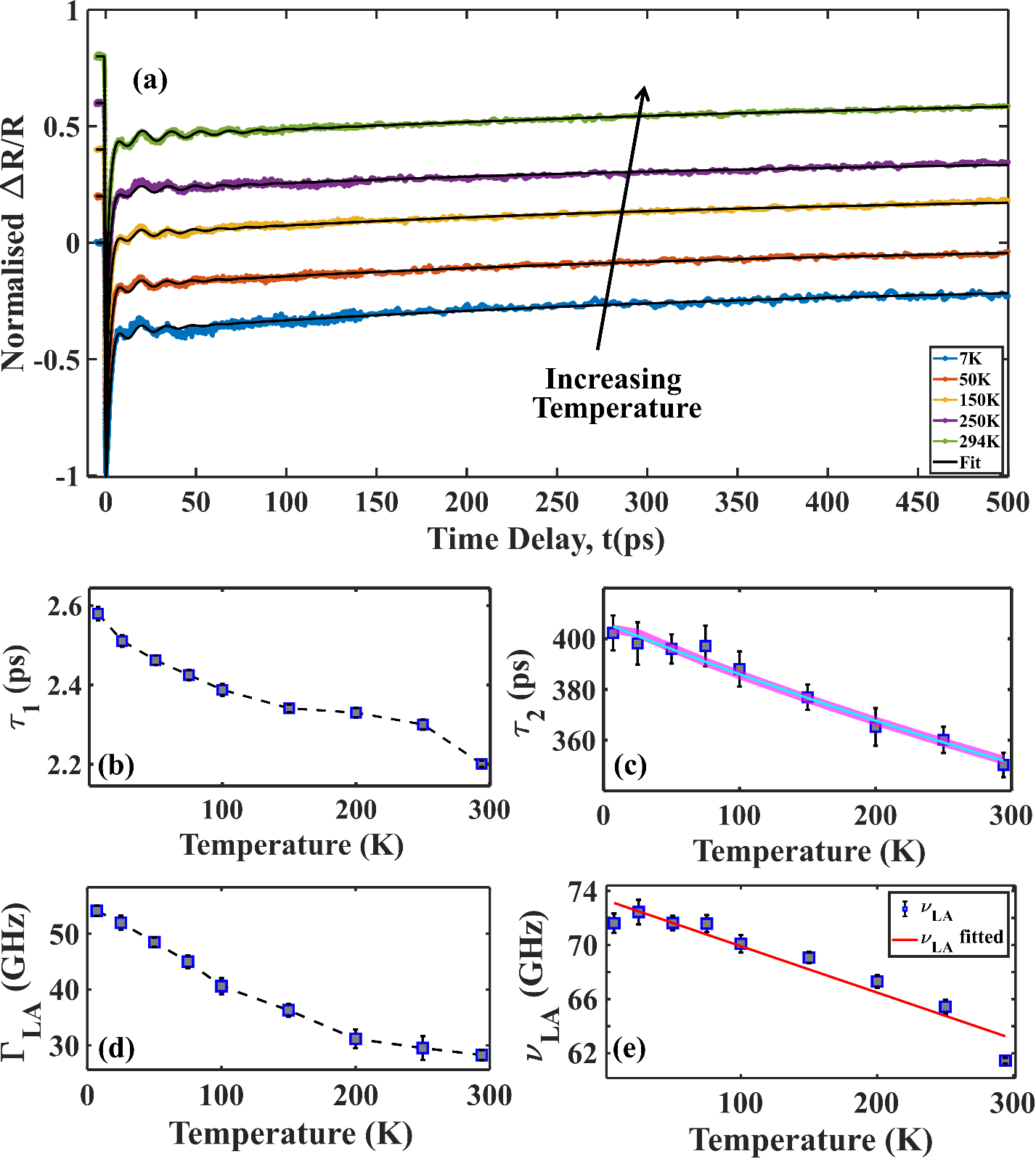}
\caption{(a) The normalised time-resolved $\Delta$R/R signal for 22 nm BSTS film at various temperatures and black solid line shows the fit. The data has been shifted vertically for clarity. (b) and (c) display the variation of $\tau_{1}$ and $\tau_{2}$ for the carriers with temperature from 7-294 K respectively. Pink and cyan line in (c) are the fit using phonon-assisted recombination model (eq. (\ref{eq_2})) with $\omega_{r}/2\pi$ = 2.0 THz and 0.073 THz respectively. (d) and (e) display the variation of the damping parameter ($\Gamma_{LA}= $1/$\tau_{LA}$) and frequency ($\nu_{LA}$) of the CAPs respectively with temperature from 7-294 K. Red solid curve is the fit using anharmonic decay model. Here, the data are represented by blue open squares and dotted lines are guide to the eye.} \label{fig_7}
\end{figure}

To understand the carrier and CAPs dynamics in BSTS further, temperature-dependent investigations were carried out from 7-294 K at fixed carrier density of 1.7 $\times$ 10$^{19}$ cm$^{-3}$ on 22 nm and 192 nm BSTS films. Fig. \ref{fig_7}(a) shows the time-resolved $\Delta$R/R signal for 22 nm BSTS film at various temperatures. The data is fitted with eq.(\ref{eq_1}) (black solid line in Fig. \ref{fig_7}(a)) and the parameters obtained are shown in Fig. \ref{fig_7}(b)-(e). We observe that $A_{0}$, $A_{1}$ and $A_{2}$ contribute around $\sim$ 8-16 $\%$, $\sim$ 70-75 $\%$ and $\sim$ 12-20 $\%$ respectively in the whole temperature range (see supplemental material \cite{Supplemental}, Fig. S10). The decay time constants $\tau_{1}$ and $\tau_{2}$, decrease by $\sim$ 15 $\%$ and $\sim$ 12 $\%$ respectively in the measured temperature range (see Fig. \ref{fig_7}(b)-(c)). The behaviour of $\tau_{1}$ versus temperature also corroborates our earlier attribution of this constant to an interplay of electron-phonon scattering and diffusion of carriers as seen in similar narrow band-gap semi-metal Te \cite{kamaraju2010large}. After thermalisation between electrons and phonons, electrons can accumulate near the conduction band minima and subsequently recombine with the holes near the valence band maxima. Since the  direct recombination process is of the order of $\sim$ nanoseconds \cite{prakash2017origin}, it cannot explain the observed temperature-dependent behaviour of $\tau_{2}$. However, phonon-assisted electron-hole recombination, in which an electron and a hole recombine through a phonon, can account for $\tau_{2}$. Previous time-resolved studies on semimetals and topological materials like Bi \cite{lopez1968electron}, WTe$_{2}$ \cite{dai2015ultrafast}, NiTe$_{2}$ \cite{cheng2022ultrafast} and MnBi$_{2}$Te$_{4}$ \cite{cheng2024magnetic} have demonstrated that the phonon-assisted electron-hole recombination (see schematic Fig. \ref{fig_6}(d)) time exhibits a significant dependence on temperature, as the phonon population is strongly temperature-dependent. At low temperatures, the electron-hole recombination time is longer due to the reduced efficiency of interband electron-phonon scattering, and a similar behaviour is experimentally observed in $\tau_{2}$ (see Fig. \ref{fig_7}(c)). In addition, the electronic structure of the BSTS allows for the possibility of flat band bending \cite{park2016possible,yilmaz2021emergent}, which favors this indirect recombination. Thus, we attribute the $\tau_{2}$ process to phonon-assisted electron-hole recombination, which can be quantitatively described by \cite{lopez1968electron}:

 \begin{equation}
     \frac{1}{\tau_{2}}= \mathscr{B} \frac{q}{sinh^{2}q}+ \frac{1}{\tau_{20}}\label{eq_2}
 \end{equation}

where $q= \hbar\omega_{r}/2k_{B}\bar{T}$ and $\omega_{r}$ is the frequency of the phonon participating in the electron-hole recombination, $\hbar$ is the reduced Planck's constant, $k_{B}$ is the Boltzmann constant and $\bar{T}$ is the temperature. $\tau_{20}$ is the temperature-independent term that is influenced by defects or impurities in the sample. The parameter "$\mathscr{B}$" signifies the density of states in the energy bands of electrons and matrix elements that govern interband scattering between electrons and holes.
Fig. \ref{fig_7}(c) shows the data and the fit with $\omega_{r}/2\pi$ = 2.00 THz (A$_{1g}^{1}$ Raman mode) and $\tau_{20}$ $\sim$ 404 ps (pink line); $\omega_{r}/2\pi$ = 0.07 THz (acoustic phonon mode) and $\tau_{20}$ $\sim$ 406 ps (cyan line). This model yields temperature-independent $\tau_{20}$ $\sim$ 404-406 ps, which is only due to impurities/defects at the lowest temperature. From our fit, it is clear that both the acoustic and optical phonons may be involved in the phonon-assisted recombination process. Thus, the decay time constant $\tau_{2}$ can be defined as: $\frac{1}{\tau_{2}}= \frac{1}{\tau_{d}}+\frac{1}{\tau_{ph}}$. Here, $\tau_{2}$ captures an effective electron-hole recombination decay channel consisting of two parallel decay processes through defects ($\tau_{d}$) and phonons ($\tau_{ph}$). In the carrier density-dependent study at room temperature, the lattice temperature remains relatively constant within $\sim$ 5K, indicating dominant defect-assisted recombination with a possible contribution from the phonon-assisted recombination process as well. However, from temperature-dependent studies, we find strong dependence of $\tau_{2}$ on the sample temperature down to 7 K indicating dominant phonon-assisted recombination process.

\begin{figure*}
\includegraphics[width=1\linewidth]{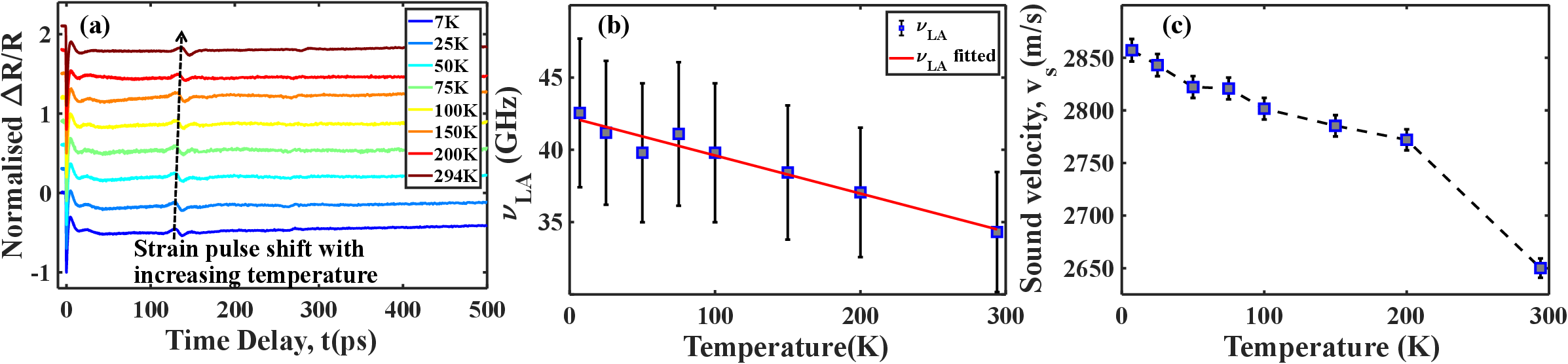}
\caption{(a) The normalised time-resolved $\Delta$R/R signal for 192 nm BSTS film at various temperatures. The data has been shifted vertically for clarity. (b) The variation of the central frequency of the strain pulse as a function of temperature. Red solid curve is the fit using anharmonic decay model. (c) The behaviour of sound velocity with increase in temperature. Here, the data are represented by blue open squares and dotted lines are guide to the eye.} \label{fig_8}
\end{figure*}

The temperature dependence of the CAP damping parameter and frequency are shown in Figs. \ref{fig_7}(d) and \ref{fig_7}(e), which suggests phonon softening with increasing temperature. The frequency of the CAP decreases from $\sim$ 71 GHz to $\sim$ 61 GHz which is nearly 14 $\%$ as the sample is heated from 7-294 K. And the damping parameter ($\Gamma_{LA}$= 1/$\tau_{LA}$) of the phonon decreases by 48 $\%$ from $\sim$ 54 GHz at 7 K to $\sim$ 28 GHz at 294K exhibiting a rather anomalous behaviour similar to previously observed for CAPs in BSTS single crystal \cite{cheng2014temperature}. The temperature dependence of the phonon mode frequency \cite{kamaraju2010large,m2024probingmagneticdimensionalcrossover,klemens1967decay,mukherjee2024magnetic,saha2009low} can be described as: $\nu_{LA}(\bar{T})$ = $\nu_{0}$ + $\Delta\nu_{anh}(\bar{T})$ where the first term $\nu_{0}$ is the frequency of the phonon at absolute zero temperature. The second term accounts for the intrinsic anharmonic contribution to phonon frequency from the decay of a phonon into two phonons (cubic anharmonicity) or three phonons (quartic anharmonicity), originating from the real part of the phonon self-energy \cite{kamaraju2010large,m2024probingmagneticdimensionalcrossover,klemens1967decay,mukherjee2024magnetic,saha2009low}. Taking into account only cubic phonon-phonon anharmonic interactions, the change in frequency is given by : $\Delta\nu_{anh}(\bar{T})$=  A$\left[1 + \frac{2}{exp(\frac{h\nu_{0}}{2k_{B}\bar{T}})-1}\right]$. Here, $h$ is the Planck's constant, $k_{B}$ is the Boltzmann constant, $\bar{T}$ is the temperature and  A is the cubic anharmonicity-induced self-energy parameter, which is negative for normal redshift. The solid red line in Fig. \ref{fig_7}(e) is the fit using the anharmonic decay model and the parameters obtained after fitting are: $\nu_{0}$ = 73.348 $\pm$ 0.595 GHz, A = -0.030 $\pm$ 0.003 GHz. 
The temperature-dependent decay of the CAP can proceed through two pathways: a phonon either decays into two phonons via cubic anharmonicity effect or, in the presence of strong electron-phonon coupling (EPC), it decays by generating an electron-hole pair \cite{kuiri2020thickness,xu2017temperature}. The damping parameter of the CAP for cubic anharmonicity effect will increase with rising temperature in contrast to the observed behaviour of $\Gamma_{LA}$ here (see Fig. \ref{fig_7}(d)). This discrepancy suggests that the EPC process should be taken into account, which is sensitive to the zero temperature joint electron–hole pair density of states at phonon energy \cite{xu2017temperature}. At low temperatures, the surface electrons near Dirac point can contribute significantly to EPC than at room temperature where thermal excitations reduce the available surface electronic states for transition \cite{xu2017temperature}. Thus, our studies here indicate that EPC plays a crucial role in the acoustic phonon decay in BSTS. 



For the 192 nm BSTS (as shown in Fig. \ref{fig_8}), we observed similar softening in the central frequency of the acoustic phonon spectrum as a function of temperature. Fig. \ref{fig_8}(a) displays the time-resolved $\Delta$R/R signal over the temperature range of 7 K to 294 K. Fig. \ref{fig_8}(b) illustrates the redshift in the frequency with increasing temperature and the solid red curve represents the fit using the cubic anharmonicity model, with the following fit parameters: $\nu_{0}$ = 42.246 $\pm$ 0.360 GHz, A = -0.013 $\pm$ 0.001 GHz. As the temporal separation between the first and second echoes of the strain pulse increases with temperature, one can estimate the variation of the sound velocity (v$_{s}$ = 2d/T$'$) as a function of temperature (see Fig. 8 (c)). We observe $\sim$ 14 $\%$ and $\sim$ 7 $\%$  reduction in sound velocity for 22 nm \cite{note_freq} and 192 nm film respectively as the sample temperature rises from 7 K to 294 K. The anomalous behaviour of the acoustic phonon damping parameter in our work here on 22 nm BSTS films requires further investigation using other techniques such as Brillouin scattering as well as comprehensive theoretical analysis. We believe that the experimental finding here will contribute in designing improved thermoelectric devices based on BSTS.

\section{Conclusion}

In conclusion, we have employed degenerate pump-probe reflection spectroscopy to study ultrafast carrier dynamics, coherent acoustic phonons, and acoustic strain pulses in BSTS topological insulator thin films of varying thicknesses. We have selected sapphire as the substrate due to its high acoustic reflection coefficient. For films with thickness  d $\gtrsim$ 2$\xi$, transient reflectivity data reveals travelling strain pulses with a single-exponential background, while for d < $\xi$, the response is dominated by CAPs and a bi-exponential electronic background with time constants  $\tau_{1}$ $\sim$ 2 ps and $\tau_{2}$ $\sim$ 260-380 ps. The strain pulse dynamics are well-described by a theoretical acoustic strain model. To further understand the physics governing photo-excited carrier, CAPs, and strain pulse dynamics, we have performed carrier density and temperature-dependent studies (7–294 K) on 22 nm and 192 nm BSTS films. The fast decay time constant ($\sim$ 2 ps) is attributed to electron-phonon scattering and carrier diffusion, while the slower decay time constant ($\sim$ 340-410 ps) is linked to defect- and phonon-assisted recombination. In the 22 nm film, CAP frequency softened with increasing temperature due to cubic anharmonicity, while the damping parameter anomalously decreased, due to acoustic phonon scattering with Dirac surface electrons. Temperature-dependent studies reveal a reduction in sound velocity with rising temperature in both 22 nm and 192 nm films. These findings provide insights into the interplay of electronic and acoustic dynamics in BSTS, offering valuable guidance for designing advanced thermoelectric devices and motivating further theoretical and experimental investigations.

\begin{acknowledgments}
The authors thank the Ministry of Education (MoE), Government of India, for funding and IISER Kolkata for the infrastructural support to carry out the research. The authors thank Prof. Debansu Chaudhuri and Debjit Biswas for the UV-Visible measurements. Anupama, Sidhanta Sahu, Poulami Ghosh, Dheerendra Singh and Sambhu G Nath thank CSIR, UGC, DST-INSPIRE, IISER Kolkata and UGC for the research fellowship, respectively. 
\end{acknowledgments}



\nocite{*}


\newpage
\onecolumngrid
\setcounter{figure}{0}
\setcounter{section}{0}
\setcounter{table}{0}
\setcounter{page}{1}
\begin{center}
\vspace{1cm}
{\Large{\bf Supplemental Material}}
\end{center}

\renewcommand{\thesection}{\Roman{section}}
\renewcommand{\thepage}{S\arabic{page}}

\setcounter{figure}{0}
\renewcommand{\figurename}{Figure}
\renewcommand{\thefigure}{S\arabic{figure}}

\setcounter{table}{0}
\renewcommand{\tablename}{Table}
\renewcommand{\thetable}{TS\arabic{table}}

\section{SAMPLE PREPARATION AND CHARACTERISATION:}

BiSbTe$_{1.5}$Se$_{1.5}$ (BSTS) films have been grown on Al$_{2}$O$_{3}$, GaAs, MgO and Si substrates using pulsed laser deposition (PLD). The target material consisted of 99.999$\%$ pure Bi, Sb, Te, and Se. The films have been grown by ablating the target with a KrF excimer laser (wavelength 248 nm) at a frequency of 2 Hz. A base pressure of 5×10$^{-6}$ mbar has been achieved before deposition, and a partial pressure of 5×10$^{-1}$ mbar of flowing Argon has been maintained throughout the deposition. High-quality film growth has been ensured by using an optimized substrate temperature of 230°C and a laser fluence of around 1 Jcm$^{-2}$. After deposition, the films were annealed at the same temperature for 20 minutes to achieve high crystallinity and reduce surface roughness. The film growth has been characterized by X-ray diffraction (XRD), Raman spectroscopy and Field emission scanning electron microscopy (FESEM).

\begin{figure}
    \centering
    \includegraphics[width=1.0\linewidth]{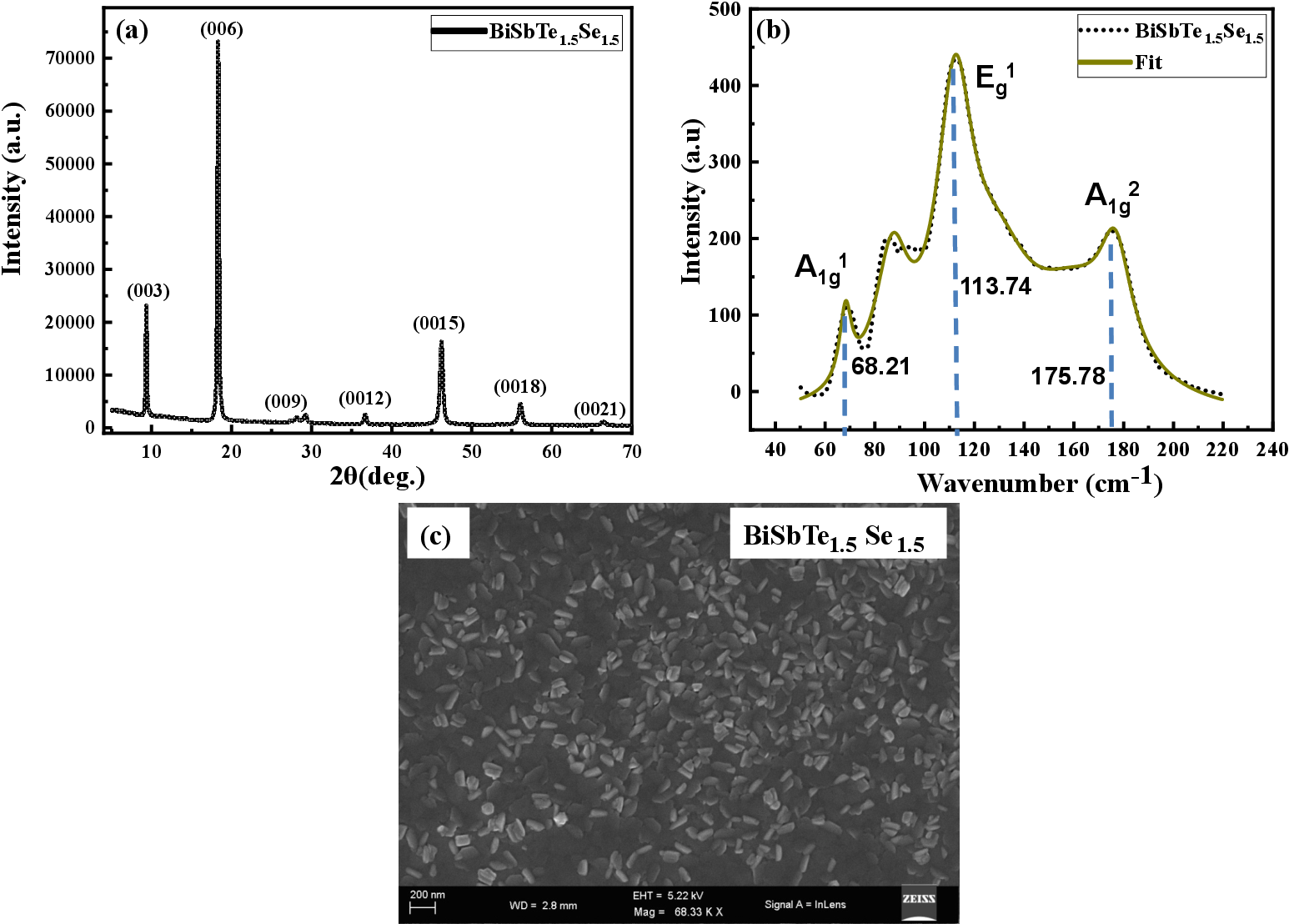}
    \caption{(a) XRD spectra (b) Raman Spectra (black dotted lines) along with multi-lorentzian fit (green solid line) (c) FESEM image for BSTS thin film.}
      \label{fig_S1_supplementary}
\end{figure}

The XRD pattern of the grown film is shown in Fig. \ref{fig_S1_supplementary}(a). From Fig. \ref{fig_S1_supplementary}(a), we find XRD peaks at 9.34° (003), 18.32° (006), 29.20° (009), 36.68° (0012), 46.21° (0015), 56.08° (0018)  and 66.35° (0021), which agrees with the values obtained previously \cite{han2018enhancement,xia2013indications}.

Raman spectrum with multiple lorentzian fit and FESEM images are shown in Fig. \ref{fig_S1_supplementary}(b) and \ref{fig_S1_supplementary}(c) respectively. Raman spectra shows peaks at 68.21 cm$^{-1}$ (A$_{1g}^{1}$), 113.74 cm$^{-1}$ (E$_{g}^{1}$) and 175.78 cm$^{-1}$ (A$_{1g}^{2}$) that agrees with references \cite{kumar2024raman,german2019phonon}. The SEM image shows uniformly distributed grains of $\sim$ 50-100 nm grain size.

\begin{figure}
    \centering
    \includegraphics[width=1.0\linewidth]{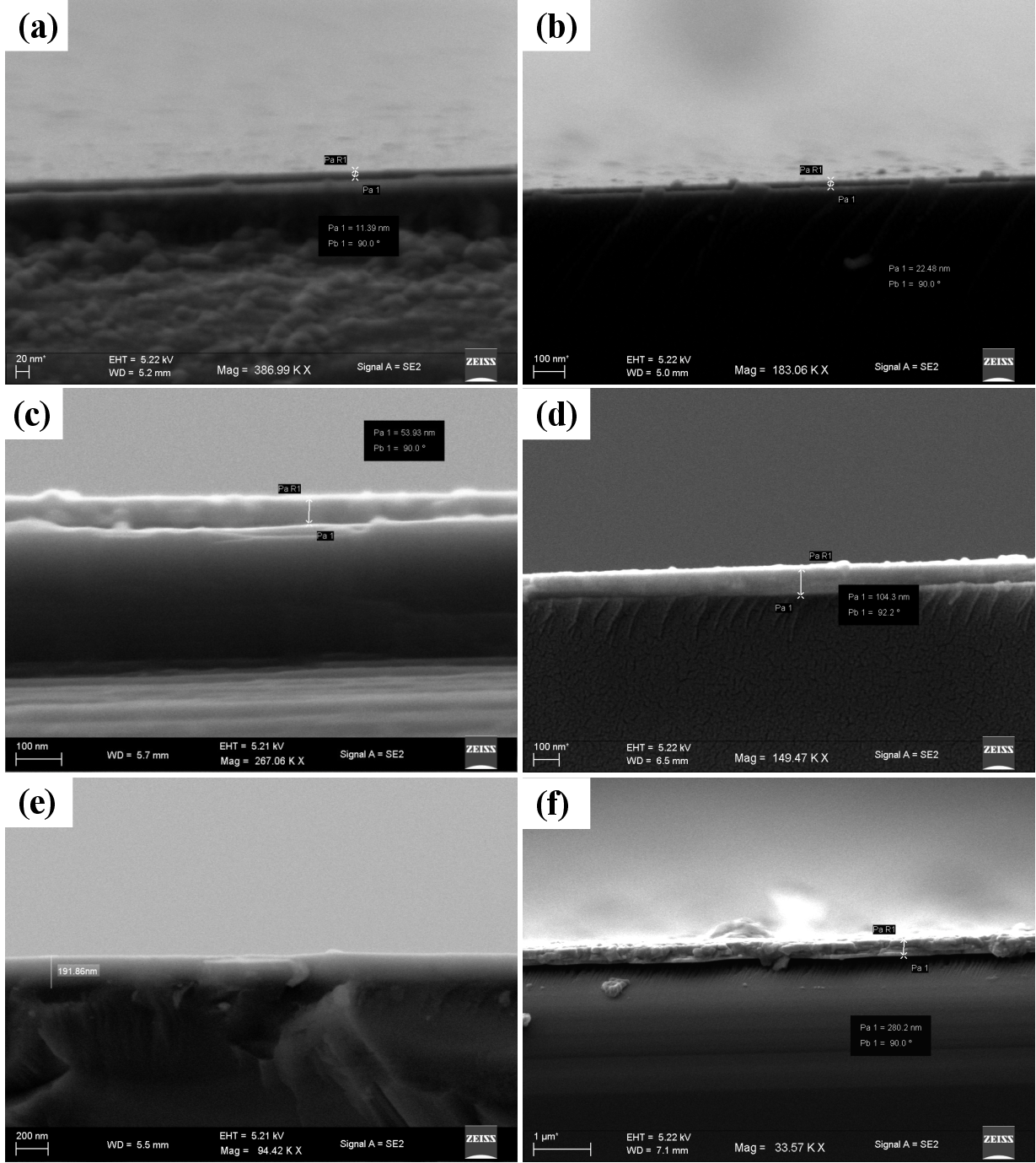}
    \caption{The Cross-sectional SEM image of BSTS films with different thickness (a) 11.4 $\pm$ 2.0 nm, (b) 22.5 $\pm$ 2.0 nm, (c) 53.9 $\pm$ 2.0 nm, (d) 104.3 $\pm$ 2.0 nm, (e) 191.9 $\pm$ 2.0 nm and (f) 280.2 $\pm$ 2.0 nm respectively.}
      \label{fig_S2_supplementary}
\end{figure}

The thickness of the BSTS films have been measured using cross-sectional SEM as shown
in Fig. \ref{fig_S2_supplementary}. The thicknesses are found to be (a) 11.4 $\pm$ 2.0 nm, (b) 22.5 $\pm$ 2.0 nm, (c) 53.9 $\pm$ 2.0 nm, (d) 104.3 $\pm$ 2.0 nm, (e) 191.9 $\pm$ 2.0 nm and (f) 280.2 $\pm$ 2.0 nm respectively.

Table \ref{tab:table1} summarises the details such as thickness, orientation, and doping of different substrates used in the experiment.

\begin{table}
\caption{\label{tab:table1}Details of different substrates used for experiment.}
\begin{ruledtabular}
\begin{tabular}{cccccc}
Substrate &	Orientation& Thickness & Polishing & Doping & Resistivity  \\ & &($\mu$m) &  & &(ohm.cm)  \\
\hline
Sapphire &	c-plane (001) &	430 &	Double sided polished &	-	&-\\
GaAs	& <100>	& 485	&Single sided polished &	Te-doped	&-\\
Si	& <111>	& 280	&Single sided polished	& Undoped	& >2000\\
MgO	& <100>	& 500	&Single sided polished	& Undoped	&-\\
\end{tabular}
\end{ruledtabular}
\end{table}

\section{EXPERIMENTAL SETUP:}

\begin{figure}
    \centering
    \includegraphics[width=0.8\linewidth]{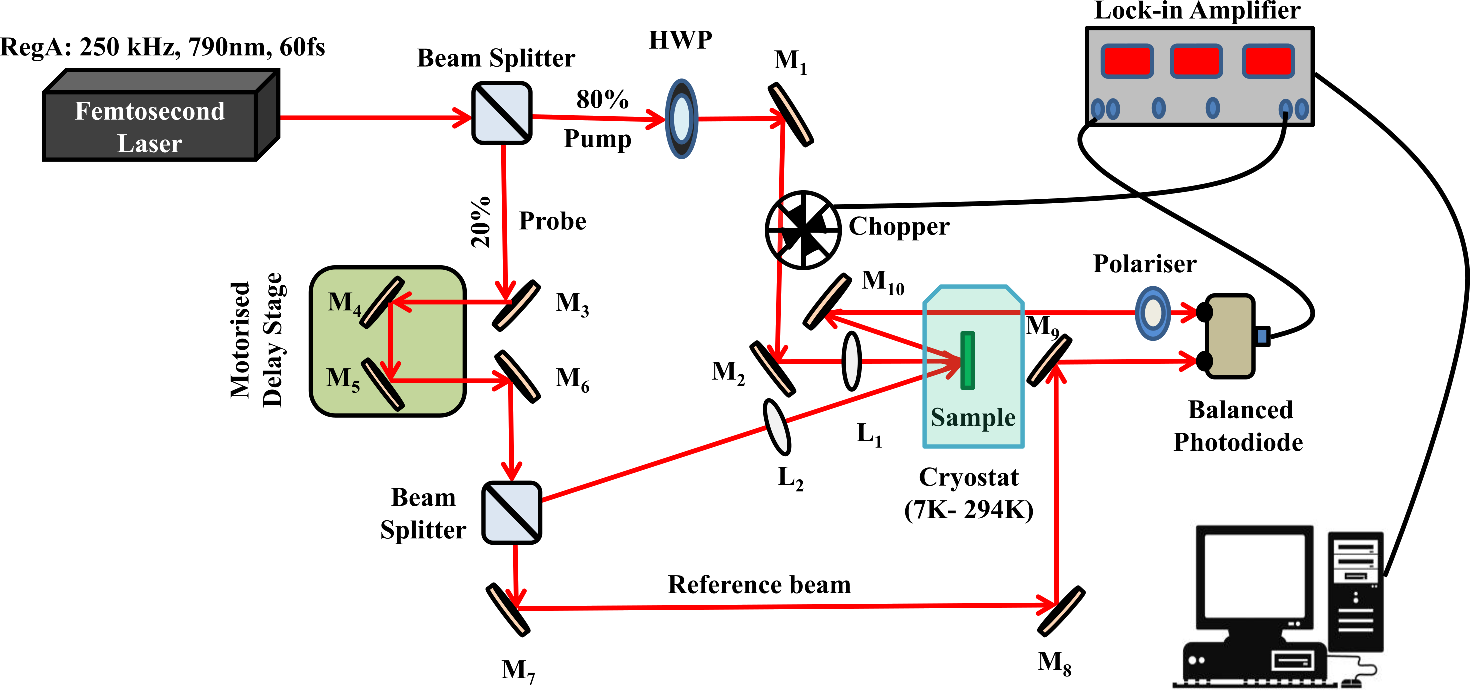}
    \caption{Schematic diagram of degenerate pump probe spectroscopy setup in reflection geometry.}
      \label{fig_S3_supplementary}
\end{figure}

Time-resolved reflectivity measurements have been done using a home-built degenerate pump-probe spectroscopy setup using a Regenerative femtosecond Amplifier (RegA 9050, Coherent Inc.) operating at a central wavelength of 790 nm with a repetition rate of 250 kHz having average output power of 1.6 W. Each pulse from this laser has a pulse width of 60 fs and pulse energy of 6.4 $\mu$J. The output pulses from the RegA are split into two parts by an 80:20 beam splitter. The more intense part is used as the pump beam and less intense part as the probe beam. A variable delay stage is employed to introduce the time delay between the pump and the probe pulses at the sample location. The sample is mounted in a cryostat (Oxford OptistatDry BLV) with a temperature range of 7 K to 294 K. The transient change in the probe beam's reflectivity induced by the pump pulse is measured using balanced photo detector. A lock-in amplifier is used to achieve the phase-sensitive detection with an optical chopper modulating the pump beam at 711 Hz. To prevent scattered pump light from reaching the detector (photo-diode), a half-wave plate (HWP) is used to rotate the pump polarisation perpendicular to the probe polarisation and a polariser is used to block the pump light while allowing the probe to pass through. The pump and probe FWHM diameters are roughly $\sim$ 49 $\mu$m and $\sim$ 39 $\mu$m, respectively. The pump fluence is varied from 11 $\mu$J/cm$^{2}$ to 174 $\mu$J/cm$^{2}$ while the probe fluence is kept constant at 6 $\mu$J/cm$^{2}$. The instrument's response function, measured by the cross-correlation between the pump and probe pulses, is $\sim$ 235 fs for our experiments. The experimental setup is shown in Fig. \ref{fig_S3_supplementary}.

\section{THEORETICAL MODEL:}
\begin{figure}
    \centering
    \includegraphics[width=0.5\linewidth]{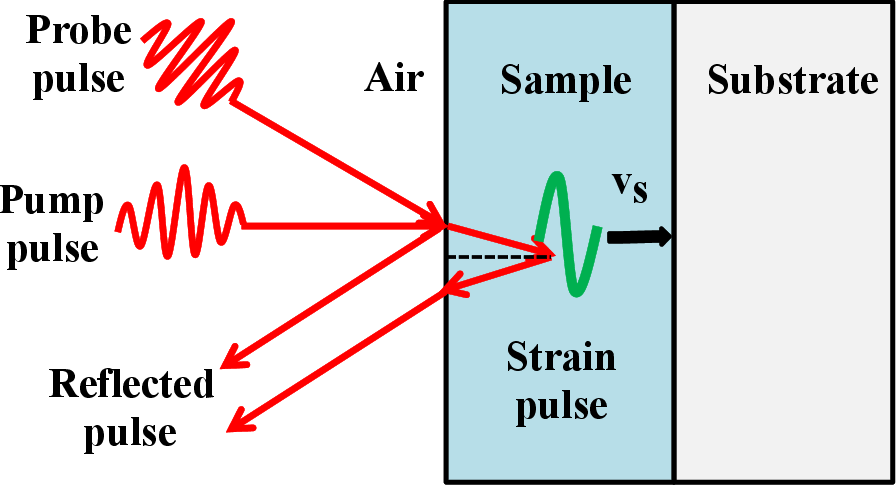}
    \caption{Schematic for the strain pulse generation and propagation }
      \label{fig_S4_supplementary}
\end{figure}

We have adopted a theoretical approach similar to those presented by Thomsen et al. \cite{thomsen1986surface} and Wu et al. \cite{wu2007femtosecond} to model the generation, propagation, and detection of travelling acoustic strain pulses (TASP) in thin films of BSTS using degenerate pump-probe spectroscopy. An intense, femtosecond laser pump pulse with the central photon energy ($\sim 1.57$ eV) above the material band gap ($\sim 0.3$ eV here in BSTS) generates both the thermoelastic ($\sigma_{l}$) and deformation potential ($\sigma_{e}$) stresses at the sample surface which launches a strain pulse propagating with the sound velocity in the material (see schematic in Fig. \ref{fig_S4_supplementary}). The stresses induced by each mechanism can be considered separately and the total stress ($\sigma_{zz}$) is just a simple sum of the electronic and thermal stresses \cite{wright2001ultrafast}:

\begin{equation}
    \sigma_{zz}(z,t)= \sigma_{l}(z,t)+\sigma_{e}(z,t)
    \label{dstress}
\end{equation}

\begin{equation}
    \sigma_{zz}(z,t)=-B\left[3\beta\Delta T(z,t)+\frac{\partial E_g}{\partial P}n(z,t)\right]
    \label{dstress}
\end{equation}

where B is the bulk modulus, $\beta$ is the linear thermal expansion coefficient, $E_g$ is the band gap, and P is the pressure; $\Delta T(z,t)$  and $n(z,t)$ are the temperature rise and photo-excited carrier density, respectively.

The wave equation for describing the generation and propagation of the stain is given by \cite{wu2007femtosecond},
\begin{equation}
    \frac{\partial^2 \eta_{zz}}{\partial t^2}-\text{v}_{s}^2\frac{\partial^2 \eta_{zz}}{\partial z^2}=\frac{1}{\rho}\frac{\partial^2 \sigma_{zz} }{\partial z^2}.
    \label{wveqn}
\end{equation}

\begin{figure}
    \centering
    \includegraphics[width=1\linewidth]{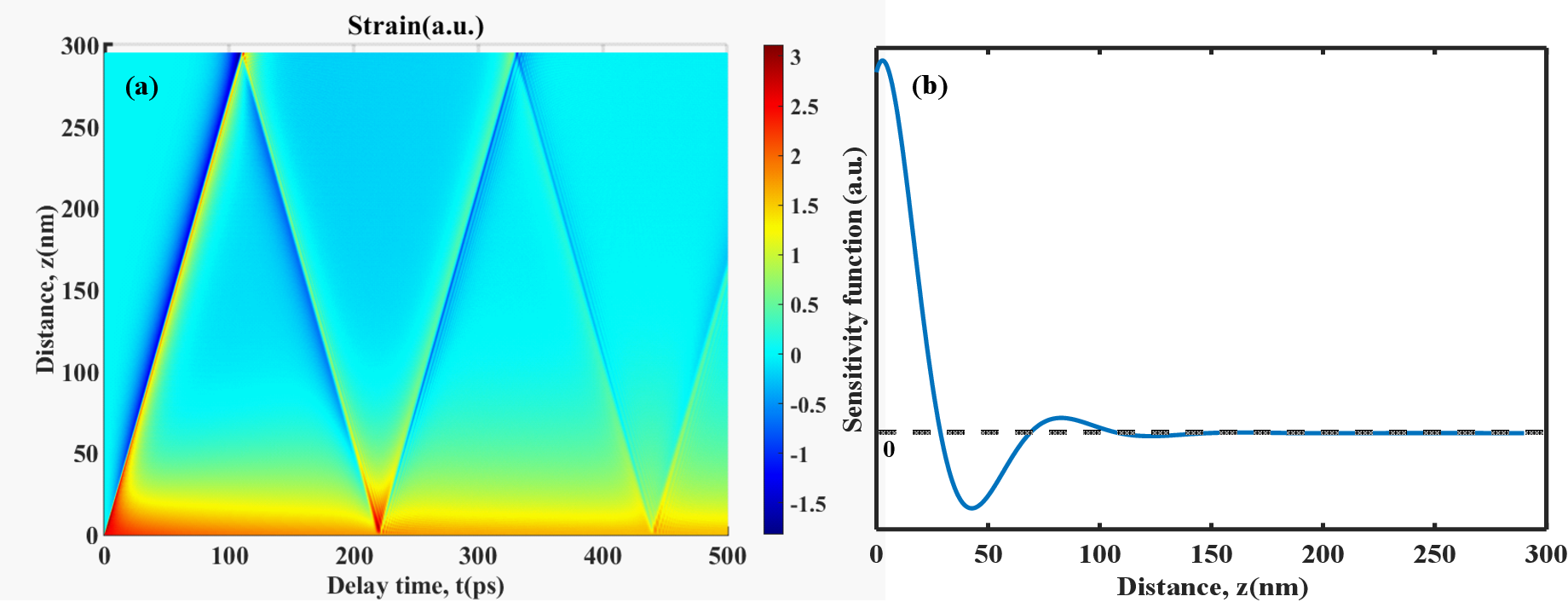}
    \caption{(a) The strain as a function of distance (z) and time delay (t) and (b) The sensitivity function as a function of z.}
      \label{fig_S5_supplementary}
\end{figure}

Here, $\rho$ is the density of the material and v$_{s}$ is the longitudinal sound velocity, which is expressed as v$_s^2=\frac{3B}{\rho}\left(\frac{1-p}{1+p}\right)$. Here, $p$ is the poisson's ratio. After putting the $n(z,t)$, $\Delta T(z,t)$ and $\sigma_{zz}$(z,t)  distributions, one can numerically solve \cite{wu2007femtosecond} the eq. (\ref{wveqn}), subject to the initial, t = 0 and elastic boundary, z = 0 conditions (eq. (\ref{eq_S4}) and eq. (\ref{strain_eq})) to obtain the strain equation.

\begin{equation}
    \eta_{zz}(z,0)=0,
    \label{eq_S4}
\end{equation}

\begin{equation}
    \eta_{zz}(0,t)=\frac{1}{\rho \text{v}_{s}^2}\left[3 B\beta \Delta T(0,t)-a_{cv}n(0,t)\right] \label{strain_eq}
\end{equation}

where, $a_{cv}=-B\frac{\partial E_g}{\partial P}$ is the relative deformation potential coupling constant, defined by the difference between the coupling constants $a_c$ and $a_v$ of the conduction and valence bands. This strain ($\eta_{zz}$ (z,t))  is combination of the strain due to thermal expansion near the surface and travelling strain propagating away from surface with the longitudinal speed of sound \cite{thomsen1986surface} (v$_{s}$)(shown in Fig. \ref{fig_S5_supplementary} (a)). It is important to note that the amplitude of strain is affected by both thermal and electronic stresses, depending on strength of their respective stresses. 

This propagating strain modifies the dielectric constant ($\epsilon = (n + i \kappa)^{2}$), which in turn changes the reflectivity ($\Delta R$) of the film. A time-delayed probe pulse measures these reflectivity changes at the surface of the film. Therefore, the change in reflectivity $\Delta R$ upto first order in strain is given as \cite{thomsen1986surface,wu2007femtosecond}: 

\begin{equation}
    \Delta R(t)= \int_{0}^{\infty}f(z)\eta_{zz}(z,t)dz,
\end{equation}
where,
\begin{equation}
    f(z)=f_0\left[\frac{\partial n}{\partial \eta_{zz}}\sin\left(\frac{4\pi n z}{\lambda}-\phi\right)+\frac{\partial \kappa}{\partial \eta_{zz}}\cos\left(\frac{4\pi n z}{\lambda}-\phi\right)\right]e^{-z/\xi},
\end{equation}
\begin{equation}
    f_0=8\frac{\omega[n^2(n^2+\kappa^2-1)^2+\kappa^2(n^2+\kappa^2+1)^2]^{1/2}}{c[(n+1)^2+\kappa^2]^2}, ~~\&
\end{equation}
\begin{equation}
    \phi=\tan^{-1}\left[\frac{\kappa(n^2+\kappa^2+1)}{n(n^2+\kappa^2-1)}\right]
\end{equation}
where $f(z)$ is the sensitivity function, $\lambda$ is the light wavelength in free space, $n$ and $\kappa$ are real and imaginary part of complex refractive index, $\xi$ is the absorption length ($\xi=c/2\omega\kappa$) and $\phi$ lies between 0 and $\pi$/2. The general form of $f(z)$ is an exponentially-damped oscillation with non zero phase at the surface i.e. at $z$ = 0 (shown in Fig. \ref{fig_S5_supplementary} (b)). Parameters used for the simulation are listed in table \ref{tab:table2}. 

\begin{table}
\centering
\caption{\label{tab:table2}Parameters used for the simulation}

\setlength{\tabcolsep}{2.5em} 
\begin{tabular}{ |lc| c|}

\hline
Parameters &	 Values \\
\hline
Refractive index, n \cite{FANG2020144822} & 5 \\
Penetration depth, $\xi$ (nm) \cite{cheng2014temperature} & 26  \\
Mass density, $\rho_{m}$ (g/cm$^{3}$) \cite{article} & 6.87\\
Longitudinal sound velocity, v$_{s}$ (m/s) & 2650\\
Bulk modulus, B (GPa) \cite{cheng2014temperature} & 37\\
Thermal expansion coeff., $\beta$ (K$^{-1}$) \cite{cheng2014temperature} & 4.9 $\times$ 10$^{-5}$ \\
Specific heat, C (JK$^{-1}$mol$^{-1}$) \cite{cheng2014temperature} & 124.4 \\
Deformation pot. for electrons and holes \cite{witting2019thermoelectric}, d$_{c}$ and d$_{v}$ (eV)& 22 and 25\\
Ambipolar diffusion constant, D$_{a}$ (cm$^{2}$s$^{-1}$) \cite{kumar2011spatially} & 500 \\
Thermal diffusion constant, D (cm$^{2}$s$^{-1}$) \cite{kumar2011spatially}  & 1.2 \\
\hline
\end{tabular}
\end{table}

The ratio \cite{thomsen1986surface} between the deformation potential stress ($\sigma_{e}$) and thermo-elastic stress ($\sigma_{l}$) can be determined by,

\begin{equation}
    \frac{\sigma_{e}}{\sigma_{l}}=\frac{C_{l}}{3\beta (E-E_{g})}\frac{dE_{g}}{dP}
\end{equation}

where $\beta $ is linear thermal expansion coeff., $C_{l}$ is specific heat and $\frac{dE_{g}}{dP}$ is the change of band gap energy with pressure. Using the same parameters as Bi$_{2}$Te$_{3}$ \cite{chen2011thermal,ovsyannikov2008giant} ($\beta$ $\sim$ 4.9 × 10$^{-5}$ K$^{-1}$, $C_{l}$ $\sim$ 124.4 JK$^{-1}$mol$^{-1}$ and $\frac{dE_{g}}{dP}$ $\sim$ 2 meV/kbar) for BSTS, we obtain, $\frac{\sigma_{e}}{\sigma_{l}}$ $\sim$ 0.065 which indicates that the thermoelastic stress has a dominant role in generating the strain.

\section{Acoustic reflection coefficient:}

\begin{figure}
    \centering
    \includegraphics[width=0.5\linewidth]{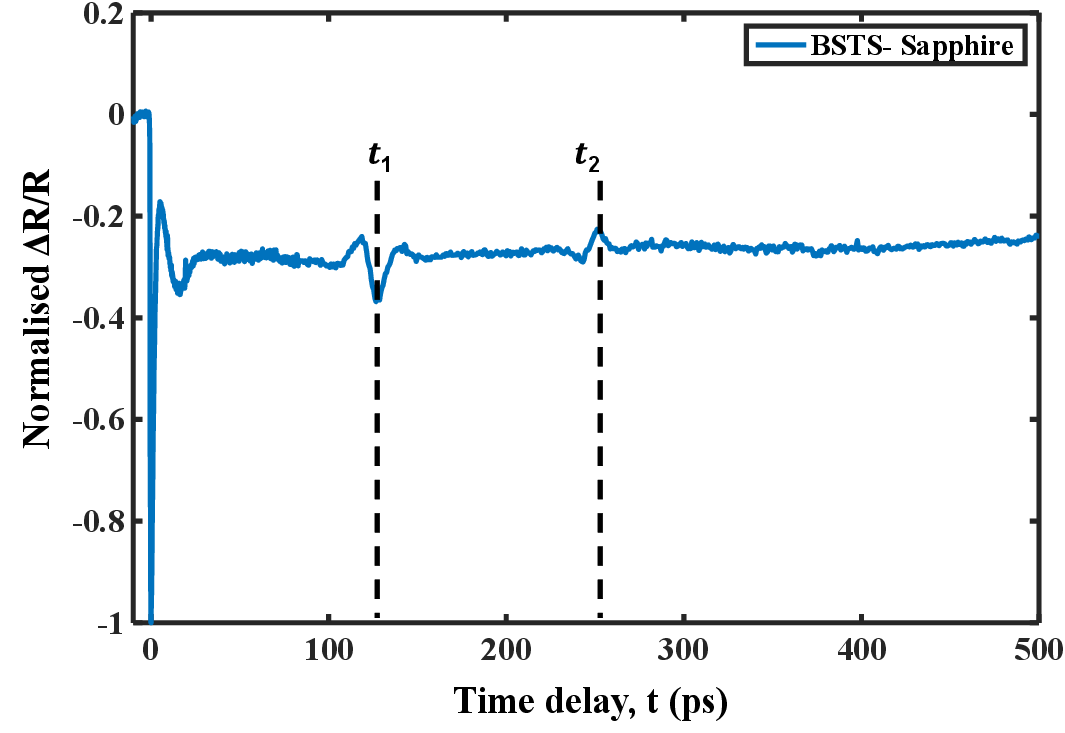}
    \caption{Time-resolved  $\Delta$R/R signal observed in BSTS thin film. Here $t_{1}$, $t_{2}$ represents the arrival time of the first and second echo of the strain pulse.}
      \label{fig_S6_supplementary}
\end{figure}

The longitudinal sound velocity in the BSTS film can be calculated by using the thickness of the film (d) and time separation between the echo pulses (T$'$= $t_{2}$- $t_{1}$, see Fig. \ref{fig_S6_supplementary} ) which is found to be 2650 m/s $\pm$ 36 m/s using the formula v$_{s}$= 2d/T$'$ in close agreement with 2600 m/s measured previously \cite{wang2010acoustic} for Bi$_{2}$Te$_{3}$.

The reflection coefficient of these acoustic strain pulses (see the insets of Fig. 1) at the submerged interface between the BSTS film and different substrates can be determined by acoustic mismatch theory \cite{thomsen1986surface}, using the formula,

\begin{equation}
    r_{a}=\frac{Z_{Substrate}-Z_{BSTS}}{Z_{Substrate}+Z_{BSTS}}
    \label{acoustic_imp}
\end{equation}

Here, r$_{a}$  is acoustic reflection coefficient, Z$_{BSTS}$ (Z$_{Substrate}$) is the acoustic impedance of BSTS (Substrate). Z$_{BSTS}$= $\rho_{BSTS}$ v$_{BSTS}$ and Z$_{Substrate}$= $\rho_{Substrate}$ v$_{Substrate}$ where $\rho_{BSTS}$ $(\rho_{Substrate})$ is the density of BSTS (Substrate) and v$_{BSTS}$ (v$_{Sapphire}$) is the longitudinal sound velocity in BSTS (Substrate). The densities of the BSTS \cite{article}, Sapphire \cite{stanton2006propagation}, GaAs \cite{kuok2000acoustic}, MgO \cite{zhang2023sound} and Si \cite{leibacher2014impedance} are 6.87 g/cm$^{3}$, 3.98 g/cm$^{3}$, 5.31 g/cm$^{3}$, 3.58 g/cm$^{3}$ and 2.33 g/cm$^{3}$ respectively. And the sound velocities for Sapphire \cite{stanton2006propagation}, GaAs \cite{kuok2000acoustic}, MgO \cite{zhang2023sound} and Si \cite{leibacher2014impedance} are $\sim$ 11250 m/s, $\sim$ 4683 m/s , $\sim$ 9000 m/s and $\sim$ 9133 m/s respectively. The calculated values using eq. (\ref{acoustic_imp}) are listed in Table \ref{tab:table3}.



\begin{table}
\caption{\label{tab:table3}List of calculated acoustic impedance for air, BSTS, sapphire, GaAs, Si and MgO; and acoustic reflection coefficient between the film and substrate.}
\begin{ruledtabular}
\begin{tabular}{lcr}
Material & Acoustic Imp.	 
&Acoustic Ref. Coeff. \\ & $(kgm^{-2}s^{-1})$ & (r$_{a}$) \\
\hline
Air	  & 413 \\
BSTS         &  1.82 × 10$^{7}$ \\
Sapphire	 & 	4.48× 10$^{7}$  &  0.42\\
GaAs		 & 	2.48× 10$^{7}$  &  0.15\\    
MgO		     &  3.22× 10$^{7}$  &  0.28\\
Si	         &  2.13× 10$^{7}$  &  0.08\\
\end{tabular}
\end{ruledtabular}
\end{table}

\section{Extraction of exponential background for all thicknesses :}

\begin{figure}
    \centering
    \includegraphics[width=0.65\linewidth]{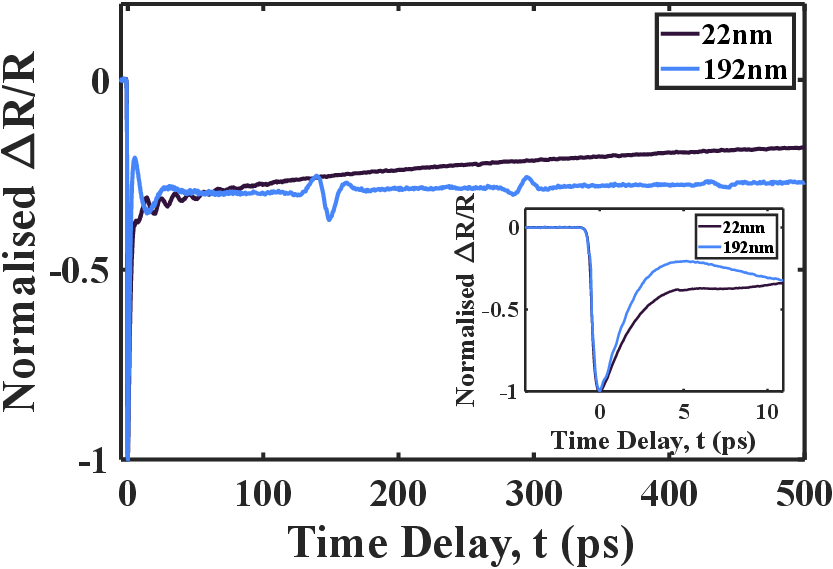}
    \caption{The time-resolved reflectivity data for 22 nm and 192 nm BSTS films. Inset shows the enlarged view of the data.}
      \label{fig_S7_supplementary}
\end{figure}

The time-resolved reflectivity data has been fitted with the bi-exponential model for 11 nm and 22 nm BSTS films, and single-exponential model for 54 nm, 104 nm, 192 nm and 280 nm films and the extracted parameters are shown in table \ref{tab:table4}. After subtracting the exponential background, the fast Fourier (FFT) intensity is plotted in insets of Fig. 2 in the main manuscript. For higher thickness, 54 nm - 280 nm, after subtracting the exponential background, time- domain data is cut and FFT is done only for the first echo of the acoustic strain pulse. The bi-exponential background for the lower thicknesses (e.g. 22 nm) is dominant as compared to that of higher thicknesses (e.g. 192 nm) where only the fast single exponential background is present. (see Fig. \ref{fig_S7_supplementary}).

\begin{table}
\centering
\caption{\label{tab:table4} Extracted decay amplitudes and decay time constants for all the BSTS films }

\setlength{\tabcolsep}{2.5em} 
\begin{tabular}{|c c c c c c|}

\hline
Thickness & A$_{0}$  & A$_{1}$ & A$_{2}$ &  $\tau_{1}$ &  $\tau_{2}$ \\ (nm) & $(\%)$  & $(\%)$& $(\%)$ & (ps) & (ps)\\
\hline
11  & 14 & 68 & 17 & 2.0 & 263 \\
22  & 14 & 73 & 13 & 2.3 & 380 \\
54  & 27 & 73 & - & 2.1 &   -  \\
104 & 14 & 86 & - & 2.5 &   -  \\
192 & 16 & 84 & - & 2.2 &   -  \\
280 & 23 & 76 & - & 2.6 &   -  \\
\hline
\end{tabular}
\end{table}

\section{TTM and MODIFIED TTM MODEL for the 22 \lowercase{nm} film :}

\begin{figure}
    \centering
    \includegraphics[width=0.65\linewidth]{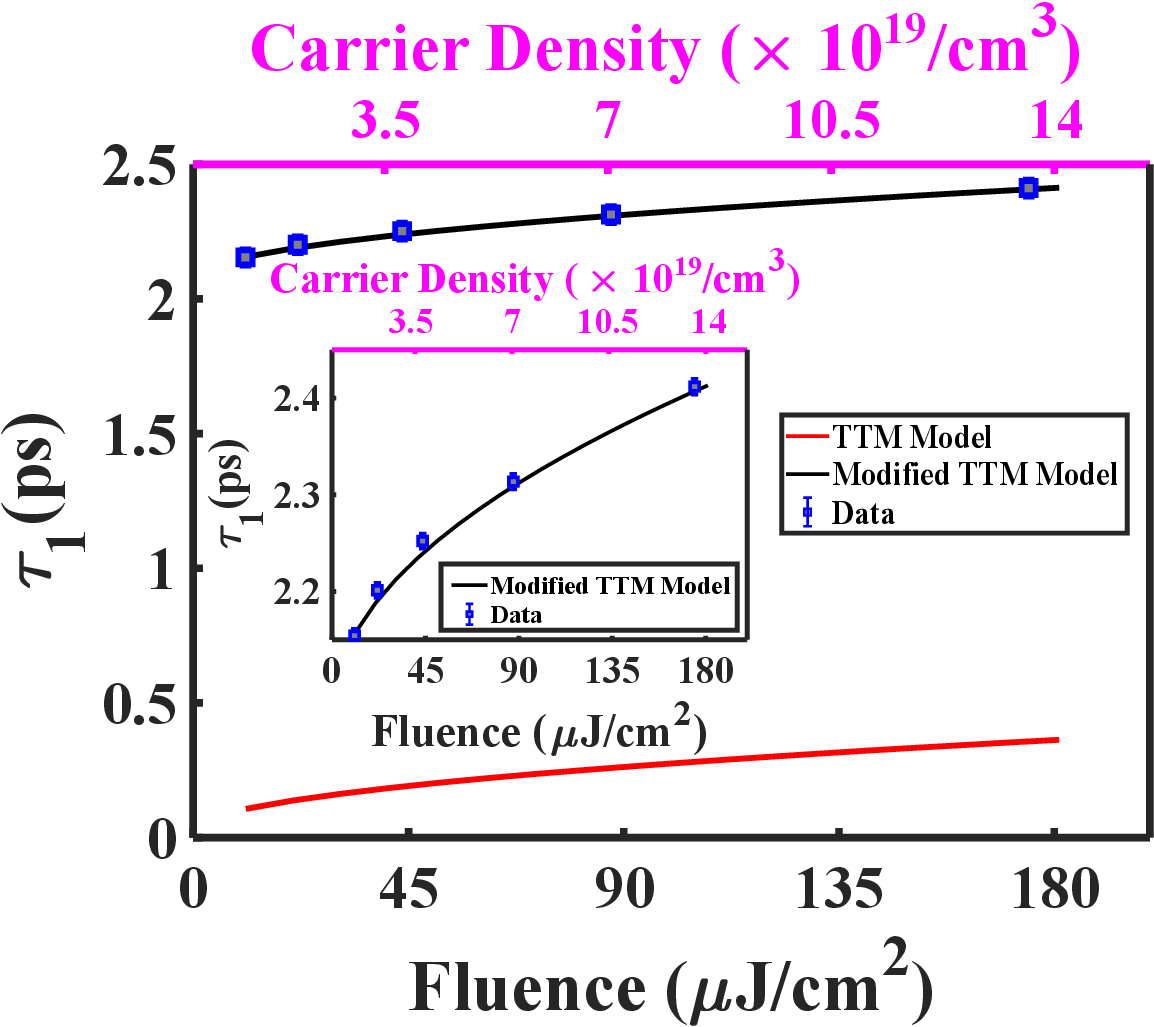}
    \caption{The variation of the fast decay constant $\tau_{1}$ with the pump fluence (the corresponding carrier densities are shown on the top axis). Here, the  blue open squares are the data, the red and black solid lines are the fit using normal  and the modified TTM \cite{groeneveld1995femtosecond}. Inset shows the enlarged view of the modified TTM fit.}
      \label{fig_S8_supplementary}
\end{figure}

Using the two-temperature model (TTM) \cite{groeneveld1995femtosecond}, the electron-phonon relaxation time is represented as:  

\begin{equation}
    \tau_{e-ph}= \frac{\gamma T_{e}^{2}-T_{l}^{2}}{2H(T_{e},T_{l})} \label{TTM_tau}
\end{equation}

where $\gamma $ is the electronic specific heat, $T_{e}$ is the electron temperature, $T_{l}$ $\approx$ 294 K is the lattice temperature and $H(T_{e},T_{l})$ is the energy transfer function and is given by,

\begin{equation}
    H(T_{e},T_{l} )=f(T_{e} )- f(T_{l} )
\end{equation}
where
\begin{equation}
     T_{e}= (T_l^{2}+\frac{2U_{l}}{\gamma})^{1/2}
\end{equation}
and
\begin{equation}
     f(\bar{T})= 4g_{\infty} \frac{\bar{T}^{5}}{\theta_{D}^{4}} \int_{0}^{\theta_{D}} \frac{x^{4}}{e^{x}-1}dx
\end{equation}			          

Here, $\theta_{D}$  represents the Debye temperature, $g_{\infty}$ is the electron-phonon coupling constant, and $U_{l}$ is the deposited energy density. The TTM model is used to explain $\tau_{1}$ for the 22 nm film at the room-temperature. As shown in Fig. \ref{fig_S8_supplementary}, the TTM (eq. (\ref{TTM_tau})) exhibits a significant deviation from the experimental data. To address this, we modified the TTM (eq. (\ref{mod_TTM})) by incorporating an additional time constant into the existing model. As seen in the inset of Fig. \ref{fig_S8_supplementary}, the modified TTM provides a much better fit to the experimental data.

\begin{equation}
    \tau_{e-ph(modified)}= \tau_{e-ph} +\tau_{0}
    \label{mod_TTM}
\end{equation}

where $\tau_{0}$ is an additional time constant. $\theta_{D}$, $ g_{\infty}$, $ U_{l}$, $\tau_{0}$ have been used as fitting parameters and the best fitting results have been obtained with $\theta_{D}$ = 156 K, $g_{\infty}$ = 3.7 $\times$ 10$^{16}$ Wm$^{-3}$K$^{-2}$, $\gamma$ = 4.8 Jm$^{-3}$K$^{-2}$ and $\tau_{0}$ = 2.05 ps.

\section{Fluence and Temperature-dependent parameters for 22 \lowercase{nm} BSTS film:}

\begin{figure}
    \centering
    \includegraphics[width=0.75\linewidth]{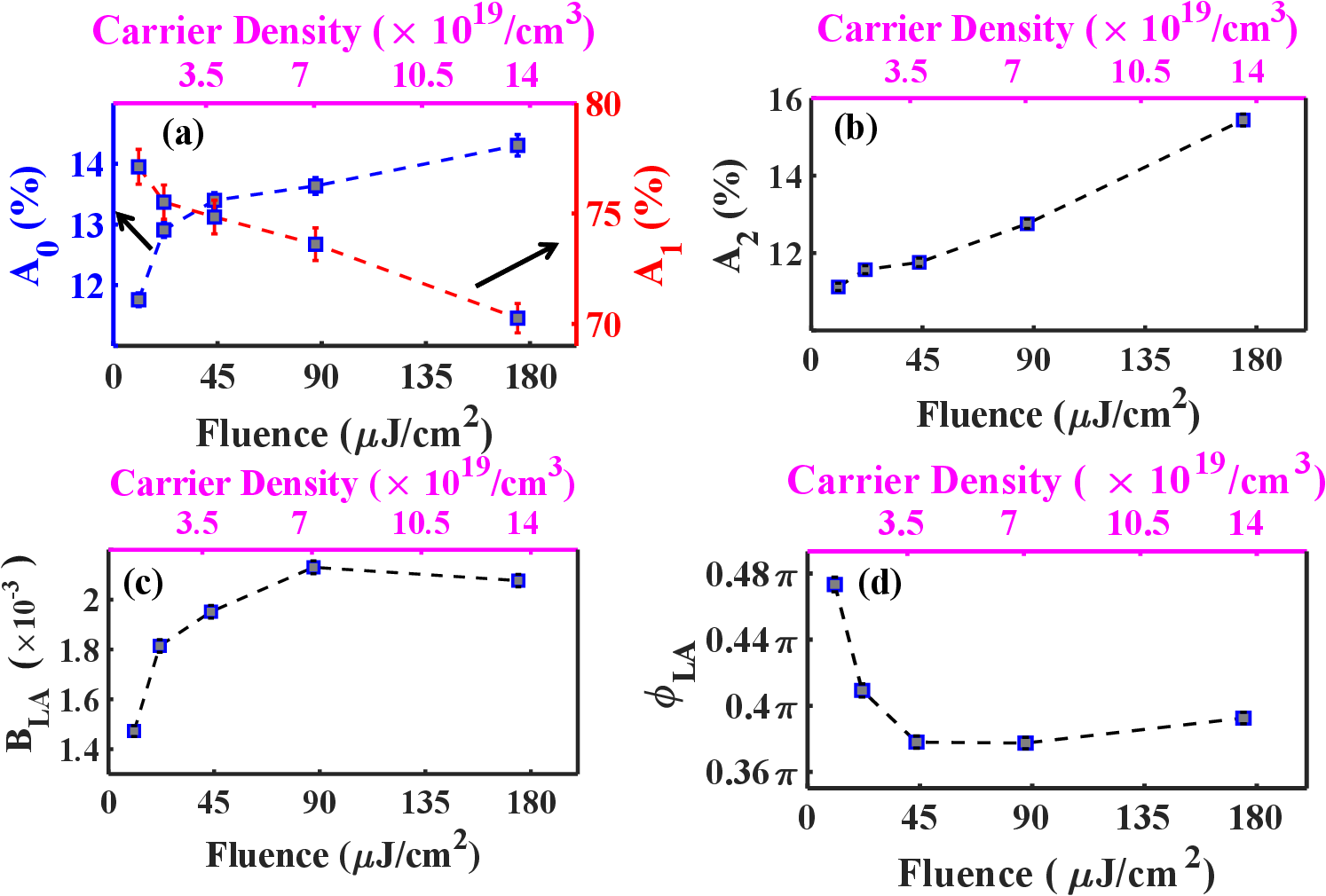}
    \caption{Pump fluence dependent variation of the fit parameters (blue open squares): (a) decay amplitudes, A$_{0}$ (left hand side), A$_{1}$ (right hand side), (b) A$_{2}$, (c) amplitude of acoustic photons, B$_{LA}$ and (d) phase $\phi_{LA}$  for 22 nm BSTS film. Top axis of all these plots display the the photo-excited carrier densities.}
      \label{fig_S9_supplementary}
\end{figure}

The behavior of the electronic decay amplitudes at various carrier densities at room temperature are shown in Fig. \ref{fig_S9_supplementary} (a)-(b). It is observed that $A_{1}$ dominates the decay process, contributing approximately 70-78$\%$, while $A_{0}$ and $A_{2}$ contribute around 12-14$\%$ and 11-15$\%$, respectively. $A_{1}$ represents the strength of the electron-phonon coupling and carrier diffusion, and it decreases by about 10$\%$ as the carrier density increases. $A_{0}$, associated with direct electron-hole recombination, increases with rising carrier density. $A_{2}$ reflects the strength of the defect-assisted and phonon-assisted recombination processes and also increases with carrier density. The variation of the CAPs amplitude, B$_{LA}$ and phase, $\phi_{LA}$ is shown in Fig. \ref{fig_S9_supplementary} (c)-(d). The amplitude of the phonon oscillation, B$_{LA}$, initially increases and then saturates at higher carrier densities, ranging from $\sim$ 1.4 $\times$ 10$^{-3}$ to 2.1 $\times$ 10$^{-3}$. Additionally, the phase of the phonon decreases from $\sim$ 0.47 $\pi$ to 0.37 $\pi$ with increasing carrier density suggesting an sine nature at low fluences, which then changes towards cosine nature at room temperature.

\begin{figure}
    \centering
    \includegraphics[width=0.75\linewidth]{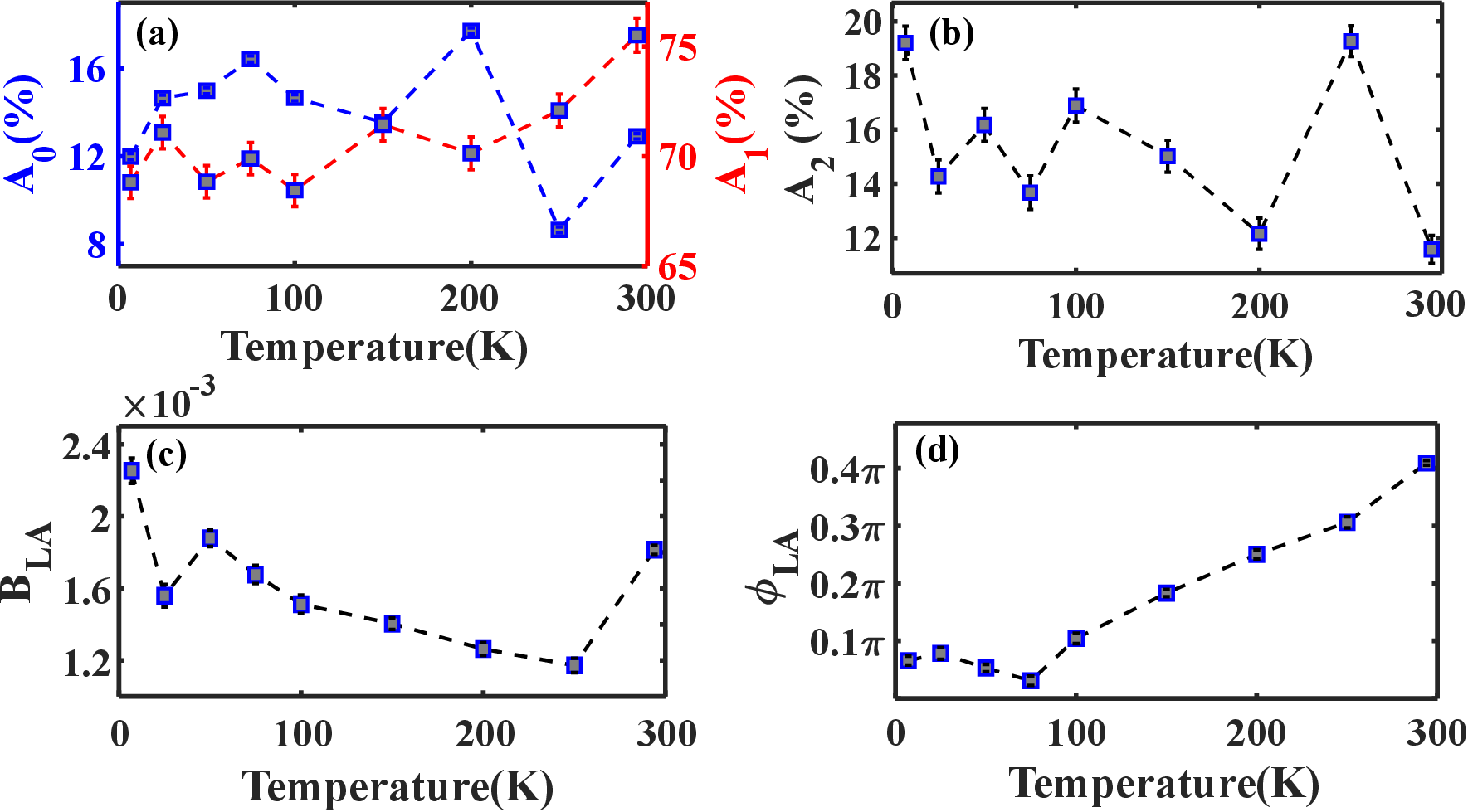}
    \caption{Temperature dependent variation of the fit parameters (blue open squares): (a) decay amplitudes, A$_{0}$ (left hand side), A$_{1}$ (right hand side), (b) A$_{2}$, (c) amplitude of acoustic photons, B$_{LA}$ and (d) phase $\phi_{LA}$  for 22 nm BSTS film. Top axis of all these plots display the the photo-excited carrier densities.}
      \label{fig_S10_supplementary}
\end{figure}

Fig. \ref{fig_S10_supplementary} (a)-(b) presents the variation of the electronic decay amplitudes, $A_{0}$, $A_{1}$, and $A_{2}$, as the sample is heated from 7 K to 294 K. $A_{0}$ and $A_{2}$ contributes approximately 12-18$\%$ and 12-20$\%$ with temperature. $A_{1}$, contributing about 70-76$\%$, increases with increase in temperature. The variation of the CAPs amplitude, B$_{LA}$ and phase, $\phi_{LA}$ with temperature is shown in Fig. \ref{fig_S10_supplementary} (c)-(d). The amplitude of the acoustic phonon decreases from $\sim$ 2.2 $\times$ 10$^{-3}$ to 1.2 $\times$ 10$^{-3}$ as temperature increases. Meanwhile, the phase of the acoustic phonon increases from approximately $\sim$ 0.06 $\pi$ to 0.41 $\pi$ with increasing temperature suggesting that the acoustic phonons here, follow a sine functional behavior at room temperature, which then changes completely to the cosine behavior at 7 K. Here the photo-excited carrier density is kept constant at $\sim$ 1.7$\times$ 10$^{19}$ cm$^{-3}$.

\section{Goodness of fit values:}

We have fitted the transient differential reflection data for the 22 nm BSTS film with the sum of bi-exponential function and oscillatory function convoluted with a Gaussian function, using eq. (1) of the main manuscript.

\begin{equation}
   \frac{\Delta R}{R}= (1-e^{-t/\tau_{r}})\times(A_{0}+\sum_{i=1}^{2} A_{i} e^{-t/\tau_{i}}
   + B_{LA} e^{-t/\tau_{LA}} cos (2\pi \nu_{LA} t + \phi_{LA}))
\end{equation}

The data and the fit are shown in Fig. 5(a) and Fig. 7(a) of the main manuscript. Goodness of fit ($\mathscr{R}^{2}$) values using eq. (\ref{R_square}) obtained after fitting the time-resolved reflectivity ($\Delta R_{exp}$) for the fluence and temperature-dependent data are listed in Table \ref{tab:table5} and \ref{tab:table6} respectively.

\begin{equation}
  \mathscr{R}^{2}= 1-\frac{\Sigma_{i}[\Delta R_{exp}(i) -\Delta R_{fit}(i)]^{2}}{ \Sigma_{i}[\Delta R_{exp}(i)- mean(\Delta R_{exp})]^{2}} \label{R_square}
\end{equation}

\begin{table}
\centering
\caption{\label{tab:table5}$\mathscr{R}^{2}$ values of the fitting for various fluences}

\setlength{\tabcolsep}{2.5em} 
\begin{tabular}{ |c c c|}

\hline
Fluence & Photo-excited Carrier Density	& $\mathscr{R}^{2}$ \\ ($\mu$J/cm$^{2}$) & ($\times$ 10$^{19}$ cm$^{-3})$ & \\
\hline
11  & 0.8   & 0.9933\\
22  & 1.7   & 0.9933\\
45  & 3.5   & 0.9956\\
87  & 6.7   & 0.9959\\
174 & 13.5  & 0.9977\\
\hline
\end{tabular}
\end{table}

\begin{table}
\centering
\caption{\label{tab:table6}$\mathscr{R}^{2}$ values of the fitting for various temperatures}

\setlength{\tabcolsep}{2.5em} 
\begin{tabular}{ |c c|}

\hline
Temperature (K) &	 $\mathscr{R}^{2}$  \\
\hline
7 & 0.9778 \\
25 & 0.9750\\
50 & 0.9926\\
75 & 0.9827\\
100&  0.9784\\
150 & 0.9924\\
200 & 0.9810\\
250 & 0.9725\\
294 & 0.9933\\

\hline
\end{tabular}
\end{table}

The carrier density can be estimated using the formula: 
N = $\frac{\alpha(1-R)F}{E_p}$ where "$\alpha$" is the absorption coefficient, "$R$" is the reflection of the probe from the sample at room temperature, "$F$" is the pump fluence and "$E_{p}$" is the central photon energy of pump pulse.

\nocite{*}

\begin{thebibliography}{100}
\bibitem{note_freq}
The reduction in longitudinal sound velocity with increasing temperature is directly proportional to the decrease in the frequency of the coherent acoustic phonon, following the relation: v$_{s}$= 2d $\nu_{LA}$.
\end{thebibliography}

\end{document}